\documentclass[aps,pra,reprint,showpacs,twocolumn,superscriptaddress,eqsecnum]{revtex4-1}
\usepackage{graphicx}
\usepackage{amsmath}
\usepackage{amssymb}
\usepackage{amsfonts}
\usepackage{bbm}
\usepackage{braket}
\usepackage{bm}
\usepackage{verbatim}
\usepackage{cancel}
\usepackage{tikz}
\usepackage{wasysym}
\usepackage[bookmarks=false,colorlinks=true,urlcolor=blue,citecolor=blue,linkcolor=blue]{hyperref}
\usepackage[normalem]{ulem}
\allowdisplaybreaks

\definecolor{darkblue}{RGB}{0,0,127}

\newcommand{\expect}[1]{\left\langle{#1}\right\rangle}
\newcommand{\dexpect}[1]{\left\langle\kern -2pt \left\langle{#1}\right\rangle\kern -2pt \right\rangle}

\begin{document}

\title{Symmetry-Protected Infinite-Temperature Quantum Memory from Subsystem Codes}

\author{Julia Wildeboer}
\email{jwildeb@iastate.edu}
\affiliation{Department of Physics and Astronomy, Iowa State University, Ames, Iowa 50011, USA}
\author{Thomas Iadecola}
\email{iadecola@iastate.edu}
\affiliation{Department of Physics and Astronomy, Iowa State University, Ames, Iowa 50011, USA}
\affiliation{Ames Laboratory, Ames, Iowa 50011, USA}
\author{Dominic J. Williamson}
\affiliation{Stanford Institute for Theoretical Physics, Stanford University, Stanford, California 94305, USA}
\email{dominic.williamson@sydney.edu.au}
\thanks{Current Address: Centre for Engineered Quantum Systems, School of Physics, University of Sydney, Sydney, New South Wales 2006,  Australia}

\date{\today}

\begin{abstract}
We study a mechanism whereby quantum information present in the initial state of a quantum many-body system can be protected for arbitrary times due to a combination of symmetry and spatial locality. 
Remarkably, the mechanism is sufficiently generic that the dynamics can be fully ergodic upon resolving the protecting symmetry and fixing the encoded quantum state, resulting in a quantum memory that persists up to infinite temperature. 
After exemplifying the mechanism in a strongly nonintegrable 
two dimensional (2D)
spin model inspired by the surface code, we find it has a natural interpretation in the language of noiseless subsystems and stabilizer subsystem codes. This interpretation yields a number of further examples, including a nonintegrable Hamiltonian with quantum memory based on the Bacon-Shor code. The lifetime of the encoded quantum information in these models is infinite provided the dynamics respects the stabilizer symmetry of the underlying subsystem code. In the presence of symmetry-violating perturbations,  we make contact with previous work leveraging the concept of prethermalization to show that the encoded quantum information can acquire a parametrically long lifetime under dynamics with an enlarged continuous symmetry group.  The prethermalization mechanism hinges on the application of external fields that are much larger than the perturbations themselves. We identify conditions on the underlying subsystem code that enable such a prethermal enhancement of the memory lifetime.
\end{abstract}

\maketitle

\section{Introduction}
Generic quantum many-body systems naturally ``forget'' their initial conditions as they evolve. Even in isolated systems undergoing unitary dynamics---where, strictly speaking, the evolution of observables always depends on the initial state---quantum information in the initial state is ``scrambled'' as entanglement between local degrees of freedom develops~\cite{Hayden07,Sekino08,Lashkari13,Landsman19}. This tendency towards scrambling is physically important, as it underlies the emergence of statistical mechanics in closed quantum systems via the eigenstate thermalization hypothesis (ETH)~\cite{Deutsch91,Srednicki94,Rigol08,D'Alessio16,Deutsch18}. However, in the context of, e.g., quantum computation, such information loss is undesirable, making the task of protecting quantum information against chaotic dynamics an important practical concern.


One of the most fascinating features of a quantum many-body system is the collective formation of a quantum error-correcting code~\cite{Shor95}. 
Quantum error correction is a key requirement for the feasibility of quantum computation in a realistic environment~\cite{Aharonov2008,Knill1998}. There is a well established subfield of research investigating the connections between topological phases of matter and the topological quantum error correcting codes they define~\cite{Kitaev2003,Bravyi98}. 
Much of the work focuses on utilizing zero- or near-zero-temperature topological phases to protect quantum information either under thermal dynamics~\cite{Dennis2001,Bacon2005a,brown2014quantum}, or via feedback and active intervention~\cite{Dennis2001}. 

More recently, the study of dynamical quantum many-body systems that naturally possess a degree of memory has gained traction. 
In this setting it is crucial to make a distinction between classical and quantum memory in a quantum system. A \textit{classical memory} is defined by the preservation of one or more classical configurations. For example, a quantum system protecting one classical bit possesses an effective ``Pauli-$Z$" operator that is conserved under the dynamics, indicating that the encoded bit never flips.
A \textit{quantum memory} is defined by the ability to form \textit{superpositions} of classical configurations that do not decohere. For example, a quantum system protecting one qubit possesses both the aforementioned effective Pauli-$Z$ operator as well as a dual ``Pauli-$X$" operator that anticommutes with it, signaling the presence of an encoded qubit with an associated Bloch sphere. 
The effective $Z$- and $X$-type operators are referred to as the logical operators of the encoded qubit.

Many examples of quantum dynamical systems with various types of memory are known.
Many-body localization (MBL)~\cite{Nandkishore15,Altman15,Abanin19} is known to approximately protect classical memory due to the presence of slow, logarithmic dephasing~\cite{Znidaric08,Bardarson12,Serbyn13a,Serbyn13b,Huse14,Swingle13,Imbrie17}. 
When MBL is combined with symmetry-protected topological order the logical operators of an encoded qubit have been shown to remain protected under symmetry-respecting dynamics~\cite{Huse13,Chandran14,Bahri15}; however, the precondition of MBL is expected to become unstable upon coupling to an external bath~\cite{Nandkishore14,Nandkishore16,Levi16,Fischer16,DeRoeck17a,Luitz17,Rubio19}. 
There have also been studies of MBL combined with topological order~\cite{Huse13,Bauer13,Wahl20}. However, for spin systems this relies on the disputed existence of MBL above one spatial dimension~\cite{DeRoeck17b,Wahl18,Gopalakrishnan19,Doggen20,Chertkov21,Decker21,hugo2020,pietracaprina2021} and, even if this were resolved, would also be expected to become unstable upon coupling to an external bath. Systems with gauge constraints or fractonic conservation laws have also been shown to lead to dynamics with classical memory related to the fragmentation of Hilbert space into disconnected subsectors~\cite{Pai19,Sala20,Khemani20,Rakovszky20,Moudgalya19,Yang20,Moudgalya21a,Moudgalya21b}. Systems with dynamical symmetries can also exhibit memory effects~\cite{Buca20,Buca21}.

Alternatively, the existence of strong zero modes (SZMs) ~\cite{Kitaev01,Fendley12} 
in a dynamical system is capable of guaranteeing a quantum memory~\cite{Fendley16,Kemp17,Else17a}. 
In this setting the SZM operator plays the role of one logical operator (say, the $Z$-type) while a global symmetry plays the role 
of a dual ($X$-type) logical operator. Symmetry-respecting dynamics  then commutes with the symmetry logical operator by definition, and  commutes with the SZM logical operator in the thermodynamic limit, again by definition.
However, in this case the quantum nature of the memory is only accessible if one considers the dynamics of states that do not respect the global symmetry. If the system is prepared in a state with a well-defined symmetry eigenvalue, then the encoded qubit is automatically restricted to be polarized along the ``$X$" axis.

There have been other efforts to protect quantum information through a combination of symmetry and topological order. 
One approach considers a 1-form symmetry-protected topological phase in three spatial dimensions that persists to nonzero temperature and supports a topological boundary phase that is self-correcting, and hence has a quantum memory, under symmetry-respecting thermal dynamics~\cite{Roberts2017,Roberts2020,Roberts2020b,Stahl2021}.
Alternatively, if the anomalous 1-form symmetry corresponding to all string operators in an Abelian topological order is strictly enforced directly in two spatial dimensions in a system with nontrivial boundary conditions (e.g., defined on a torus), any local symmetry-respecting dynamics must automatically support a quantum memory, as the nonlocal logical string operators around nontrivial cycles commute with the dynamics~\cite{Bombin2009,Kargarian2010,Bombin10,Suchara11,Bravyi2013}. 
Another approach that has experienced an explosion of interest is quantum dynamics involving hybrid ``unitary-projective'' quantum circuits in which unitary gates are combined with projective measurements~\cite{Li18,Skinner19,Li19,Chan19}. 
This has lead to the discovery of systems with both random~\cite{Gullans20,Choi20,Fan21,Fidkowski21} and topologically structured~\cite{Sang21,Lavasani21a,Ippoliti2021,Lavasani21b} quantum memory. 

Here we take inspiration from several directions of previous work and consider quantum dynamics that is defined by a combination of symmetry and locality. 
The main example we consider is a family of Hamiltonians with finite gauge symmetry in two spatial dimensions whose dynamics are shown to act as an exact quantum memory for infinite time.
In previous work, approximate quantum memories have been obtained by mechanisms that tend to suppress ergodicity of the quantum dynamics. For example, classical and quantum memories relying on MBL require the complete breakdown of ergodicity in order to operate. Other studies of approximate quantum memories rely on perturbing around an exactly solvable point with simple dynamics~\cite{Fendley16,Else17a,Kemp17} and/or considering a solvable system at finite temperature~\cite{Else17a}.
The main example considered in this work differs substantially in that it provides an exact quantum memory even in a system which otherwise exhibits \textit{fully ergodic} dynamics---that is, the family of Hamiltonians we consider supports a quantum memory even at \textit{infinite temperature}.
This infinite-temperature memory arises due to an exact twofold degeneracy of every eigenstate in the spectrum of the Hamiltonian, which implies that the memory persists under dynamics from any initial state.
Remarkably, only a global $\mathbb Z_2\times\mathbb Z_2$ symmetry is necessary to guarantee the existence of the quantum memory under locally generated dynamics. 
To study the effect of small perturbations that break these symmetries, we leverage previous work on prethermalization~\cite{Abanin17,Else17a,Else20} showing that the protecting symmetry can in principle persist as an approximate emergent symmetry for (stretched) exponentially long times in the presence of appropriate external fields. To do so, we embed the protecting $\mathbb Z_2\times\mathbb Z_2$ symmetry into a larger U(1)$\times$U(1) symmetry group, yielding a closely related class of Hamiltonians to which the prethermalization arguments can be applied.

We find that our construction is naturally understood within the formalism of stabilizer subsystem codes~\cite{Poulin2005,Bacon2005a}, which leads to a number of generalizations and further examples. This formalism can be applied not only to Hamiltonian dynamics, but also to arbitrary unitary, unitary-projective, and open-system dynamics with the appropriate symmetry.
We also consider conditions on the underlying stabilizer subsystem code that enable the application of prethermalization arguments~\cite{Else17a,Else20} that ensure a parametrically long-lived quantum memory under dynamics generated by a symmetric Hamiltonian with symmetry-breaking perturbations that are small compared to a set of applied fields. In particular, we find that it is sufficient for the underlying code to be a \textit{topological} subsystem code~\cite{Bombin10} that contains a nontrivial \textit{global} error-detecting subsystem code with no local dressed logical operators. This condition is also necessary in the sense that a Hamiltonian suitable for the application of prethermalization arguments along the lines of Refs.~\cite{Else17a,Else20} defines a topological stabilizer group.

The paper is laid out as follows. In Section~\ref{sec:Z2code} we show that the above-mentioned family of Hamiltonians based on a $\mathbb Z_2$ gauge theory with global symmetries supports an infinite temperature quantum memory. We also provide explicit numerical examples showing that this family of Hamiltonians is quantum-chaotic. We then define a closely related class of models that we argue based on previous work~\cite{Else17a,Else20} should exhibit an approximate quantum memory for (stretched) exponentially long times in the presence of symmetry-breaking perturbations. We also present numerical results investigating the effects of perturbations in these models. In Section~\ref{sec:SubsystemCodes}, we consider more general criteria under which stabilizer subsystem codes define dynamical quantum memories similar to the one constructed in our main example. Based on this discussion, we also consider further examples of such models, including one to which the prethermalization arguments of Refs.~\cite{Abanin17,Else17a,Else20} cannot be applied. 
In Section~\ref{sec:conclusion} we draw our conclusions and discuss potential future directions.


\section{Main example: 2D $\mathbb Z_2$ gauge theory with global symmetries} \label{sec:Z2code}
\subsection{Setup and global symmetry}
\label{sec:Setup}
\begin{figure}[t]
\includegraphics[width=1.00\columnwidth]{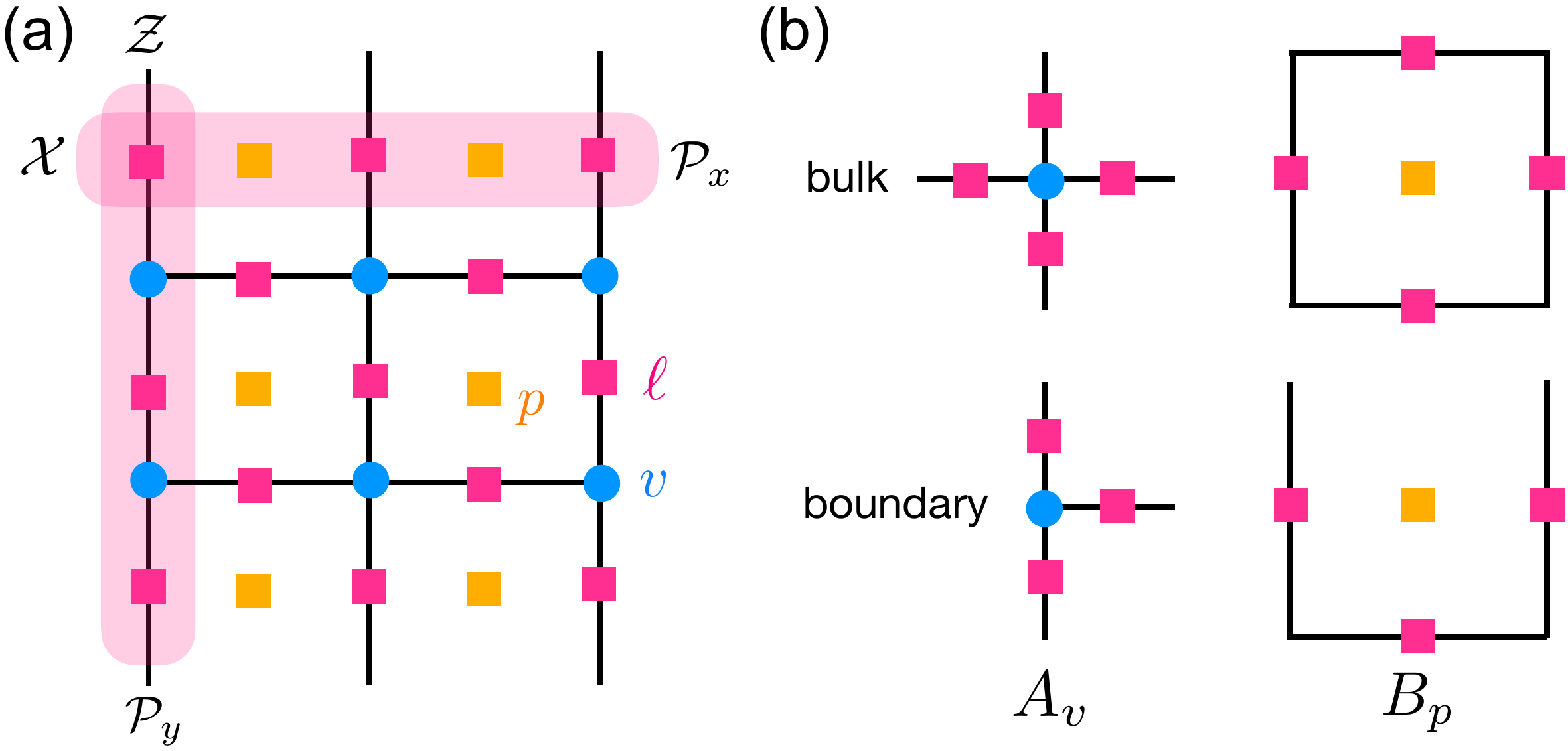}
\caption{
{\it Depiction of the main example qubit system.} 
          (a) The square lattice of size $(L_{x},L_{y}) = (2,3)$ with ``gauge'' qubits on its links $\ell$ (magenta squares), and ``matter'' qubits on its vertices $v$ (blue circles) and on its plaquettes $p$ (yellow squares). 
          Also shown 
            are the  
            logical-$X$ operator $\mathcal X = \prod_{\ell\in\mathcal P_x} X_\ell$ 
            and  the logical-$Z$ operator $\mathcal Z = \prod_{\ell\in\mathcal P_y} Z_\ell$ 
            each acting on a specific subset of links, i.e. for the logical-$X$ operator $\mathcal X$ 
            this subset consists of all dangling links on one of the rough boundaries as indicated by the (horizontal) 
            path $\mathcal P_x$, while for the logical-$Z$ operator $\mathcal Z$ the respective subset is made up 
            of all links on one of the smooth boundaries, i.e. the vertical path $\mathcal P_y$.  
        (b) The operators $A_v$ and $B_p$ acting on a vertex and a plaquette in the bulk and at the boundary, respectively.
}
\label{fig:lattice}
\end{figure}

We consider a square lattice with ``gauge" qubits on its links $\ell$ and ``matter" qubits on its vertices $v$ and on its plaquettes $p$ (see Fig.~\ref{fig:lattice}). The total Hilbert space is therefore $\bigotimes_{\ell, v, p}\mathbb{C}^2$. We refer to the full collection of gauge and matter qubits as ``physical qubits," to distinguish them from the logical qubit that is introduced later. Pauli operators for all physical qubits are denoted by $X_r,Y_r,Z_r$, with $r=\ell,v,p$ specifying the position of the qubit and 
\begin{align}
X_{r} = \begin{bmatrix}
0 & 1 \\
1 & 0 
\end{bmatrix}
\,\,\,\,\,\,\,
Y_{r} = \begin{bmatrix}
0 & -i \\
i & \,\,\,\,\,0 
\end{bmatrix}
\,\,\,\,\,\,\,
Z_{r} = \begin{bmatrix}
1 & \,\,\,\,\,0 \\
0 & -1 
\end{bmatrix}.
\end{align}
We employ boundary conditions on the square lattice inspired by the surface code~\cite{Kitaev2003,Bravyi98,Freedman01,Fowler12}, taking ``rough'' boundaries with dangling links on the top and bottom and ``smooth'' boundaries without dangling links on the left and right sides.

We define a subspace of the total Hilbert space spanned by gauge-invariant states, namely
\begin{align}
\label{eq:HilbertSpace}
    \mathcal{H}=\{\ket{\psi}\mid A_v\ket{\psi}=B_p\ket{\psi}=\ket{\psi}\},
\end{align}
with the mutually commuting gauge transformation generators $A_v,B_p$ defined as
\begin{align}
\label{eq:GaugeGenerators}
    A_v = X_v\prod_{\ell\in\ell_v}X_\ell,\indent B_p = Z_p\prod_{\ell\in\ell_p}Z_\ell,
\end{align}
with $\ell_p$ the set of links touching the plaquette $p$ and $\ell_v$ the set of links touching the vertex $v$. The sets $\ell_v$ and $\ell_p$ each contain four links for $v,p$ in the bulk of the lattice and three links for $v,p$ on the boundary.  The gauge invariance condition in Eq.~\eqref{eq:HilbertSpace} ensures that electric and magnetic excitations of the gauge spins are paired with vertex and plaquette excitations of the matter spins, respectively. For example, any state $\ket{\psi}\in\mathcal H$ containing an electric gauge excitation on vertex $v$---i.e., that satisfies $\prod_{\ell\in\ell_v}X_\ell\ket{\psi}=-\ket{\psi}$---must also satisfy $X_v\ket{\psi}=-\ket{\psi}$.

We further decompose the Hilbert space $\mathcal{H}$ into ``topological symmetry" sectors associated with eigenvalues of the global spin and phase flip operators
\begin{align}
\label{eq:TopoSymm}
    S_X=\prod_v X_v,&& S_Z = \prod_p Z_p,
\end{align}
which commute with each other and with the gauge generators.  These operators count the parity of the number of vertex and plaquette matter excitations, respectively, and enforcing these topological symmetries therefore breaks $\mathcal H$ up into four disconnected symmetry sectors that we label with indices $(\pm,\pm)$.  Note that these topological symmetry constraints are only nontrivial in the case of open boundary conditions; for periodic boundary conditions, $S_X=\prod_v A_v$ and $S_Z=\prod_p B_p$, so that only the $(+,+)$ sector is consistent with the gauge constraints.

\subsection{Symmetric and non-symmetric local operators}
We now consider the set of gauge-invariant local operators 
\begin{table*}[t!]
    \centering
    \begin{tabular}{|c|c|c|}
    \hline
    Terminology  & Sec.~\ref{sec:Z2code} & Sec.~\ref{sec:SubsystemCodes}  \\
         \hline\hline
    Gauge group & $G$ [as defined by Eqs.~\eqref{eq:SymmetricOps}] & $G$
    \\
    \hline
    Stabilizer group & $S=\langle S_X,S_Z,\{A_v\},\{B_p\}\rangle$ & $S = Z(G)$
    \\
    \hline
    Bare logical group & $L=\langle{\mathcal X,\mathcal Z, S}\rangle/S$ & $L=C(G)/S$
    \\
    \hline
    Dressed logical group&
    $\braket{\mathcal X, \mathcal Z,G}/G$
    &
    $C(S)/G$
    \\
    \hline
    \end{tabular}
    \caption{\textit{Summary of stabilizer subsystem code concepts.} This table collects the key concepts of subsystem codes---namely the gauge, stabilizer, and logical groups---and points out how these concepts are instantiated in Secs.~\ref{sec:Z2code} and \ref{sec:SubsystemCodes}. Here, $\expect{\cdot}$ denotes the multiplicative group generated by the enclosed set of operators, and $Z(G)$ and $C(G)$ denote the center and centralizer, respectively, of the group $G$.}
    \label{tab:SSC Table}
\end{table*}

acting on the Hilbert space $\mathcal{H}$.  In the bulk, such operators either reside on a vertex $v$ or plaquette $p$, or include a link $\ell(v_1,v_2)$ connecting nearest-neighbor vertices $v_{1,2}$ or a link $\ell(p_1,p_2)$ separating nearest-neighbor plaquettes $p_{1,2}$. These operators are generated by 
\begin{subequations}
\label{eq:SymmetricOps}
\begin{equation}
\label{eq:SymmetricOpsGen}
    X_v,\ Z_p,\ Z_{v_1}Z_{\ell(v_1,v_2)}Z_{v_2},\ X_{p_1}X_{\ell(p_1,p_2)}X_{p_2} , 
\end{equation}
products of which include, e.g., 
\begin{equation}
    Y_{v_1}Z_{\ell(v_1,v_2)}Y_{v_2},\ Y_{p_1}X_{\ell(p_1,p_2)}Y_{p_2}\, \dots.
\end{equation}
\end{subequations}
We remark that all of these bulk operators commute with the topological symmetry generators~\eqref{eq:TopoSymm}. 
Gauge invariant operators that fail to commute with the topological symmetries are generated by terms along the boundary of the system, 
up to multiplication by the symmetric bulk terms described above. 
For each dangling link $\ell$ along a rough boundary, with adjacent vertex $v(\ell)$, the operator
\begin{subequations}
\label{eq:NonSymmetricOps}
\begin{align}
    Z_{\ell}Z_{v(\ell)},
\end{align}
commutes with $A_v$ but anticommutes with $S_X$, and generates further terms such as $ Z_{\ell}Y_{v(\ell)}$, etc. 
For each link $\ell$ along a smooth boundary, with adjacent plaquette $p$, the operator  
\begin{align}
     X_{\ell}X_{p(\ell)}, 
\end{align}
\end{subequations} 
commutes with $B_p$ but anticommutes with $S_Z$, and generates additional terms such as $ X_{\ell}Y_{p(\ell)}$, etc.
We denote by $G$ the set of gauge-invariant local operators, including \eqref{eq:SymmetricOps}, that commute with the topological symmetry generators $S_X,S_Z$, and by $\overline G$ the set of gauge-invariant local operators, including \eqref{eq:NonSymmetricOps}, that do not commute with $S_X,S_Z$.

\subsection{Encoded logical qubit and connection to subsystem codes}
\label{sec: Encoded logical qubit and connection to subsystem codes}

We now show that the Hilbert space $\mathcal H$ supports a single encoded logical qubit when the topological symmetries~\eqref{eq:TopoSymm} are enforced. To do this, we identify a conjugate pair of nonlocal logical operators.  The logical-$X$ operator is defined as
\begin{subequations}
\label{eq:BareLogicals}
\begin{align}
    \mathcal X = \prod_{\ell\in\mathcal P_x} X_\ell,
\end{align}
while the logical-$Z$ operator is defined as
\begin{align}
    \mathcal Z = \prod_{\ell\in\mathcal P_y} Z_\ell.
\end{align}
\end{subequations}
Here, $\mathcal P_x$ is a path including all dangling links on one of the rough boundaries, while $\mathcal P_y$ is a path including all links on one of the smooth boundaries (see Fig.~\ref{fig:lattice}).
Note that these operators anticommute with each other and commute with $A_v,B_p$ and $S_X,S_Z$, as expected.  Moreover, from the definitions of the generators \eqref{eq:SymmetricOpsGen} of $G$ and \eqref{eq:NonSymmetricOps} of $\overline{G}$, we see that $\mathcal X$ and $\mathcal Z$ commute with all symmetric gauge-invariant operators in $G$, while at least one of $\mathcal X$ and $\mathcal Z$ (up to multiplication by $S_X$ and/or $S_Z$) anticommute with any non-symmetric gauge-invariant operator from $\overline{G}$. Thus, in the presence of the topological symmetries~\eqref{eq:TopoSymm}, we can decompose an arbitrary state $\ket{\psi}\in \mathcal H$ within a symmetry sector labeled by eigenvalues of $S_X$ and $S_Z$ as
\begin{align}
\label{eq:psi}
    \ket{\psi}=\ket{\psi_{\rm L}}\otimes\ket{\psi'},
\end{align} 
where operators in $G$ act nontrivally only on $\ket{\psi'}$ and where $\ket{\psi_{\rm L}}$ is the state of the logical qubit. $\ket{\psi_{\rm L}}$ is specified by the expectation values $\braket{\mathcal X},\braket{\mathcal Y},$ and $\braket{\mathcal Z}$, where $\mathcal Y =i\mathcal X\mathcal Z$. As an aside, we note that there is also an analog of this setup that encodes more than one logical qubit; we briefly discuss this setup in Appendix~\ref{sec: Encoding multiple qubits}.

The local generators we have introduced above in fact define a \textit{stabilizer subsystem code}~\cite{Poulin2005,Bacon2005a}, whose stabilizer group is generated by the gauge transformation generators $A_v,B_p$ and the topological symmetry operators $S_{X},S_{Z}$, and whose encoded qubit is $\ket{\psi_{\rm L}}$. (For a precise definition of stabilizer subsystem codes, we refer the reader to Sec.~\ref{sec:SubsystemCodes}; the present discussion is intended to be a less formal complement to that treatment. The relationships between concepts discussed in this Section and in Sec.~\ref{sec:SubsystemCodes} are summarized in Table~\ref{tab:SSC Table}.) 
This subsystem code is not particularly attractive from the point of view of practical quantum error correction, as the code distance is independent of the number $n$ of physical qubits. To see this, we note that there exist \textit{nonlocal} but few-body operators that commute with all stabilizers but anticommute with the nonlocal logical operators. An example of such an operator is
\begin{align}
\label{eq:Z2DressedLogical}
    X_{\ell_{\rm TL}}X_{v(\ell_{\rm TL})}X_{v(\ell_{\rm TR})}X_{\ell_{\rm TR}},
\end{align}
where $\ell_{\rm TL}$ and $\ell_{\rm TR}$ are the dangling links at the top-left and top-right corners of the lattice.
This operator commutes with the gauge generators $A_v,B_p$ and the topological symmetry generators \eqref{eq:TopoSymm} but anticommutes with $\mathcal Z$, and thereby generates an undectected logical error.  
In the language of subsystem codes, these are known as ``dressed logical operators'', as they can be realized as products of the ``bare logical operators'' $\mathcal X,\mathcal Z$ and operators from the set $G$ of symmetric gauge-invariant local operators~\cite{Poulin2005}. 
The operators from $G$ are, somewhat confusingly, known as ``gauge'' operators in the subsystem code literature. We use quotations to make clear that this is distinct from the gauge symmetries present in our model, which correspond to \textit{stabilizers} of the subsystem code. 

Although dressed logical operators like \eqref{eq:Z2DressedLogical} are undesirable from the viewpoint of quantum error correction, they are less detrimental in the context of locally generated quantum dynamics. To see this, imagine preparing an initial state with an arbitrary encoded state $\ket{\psi_{\rm L}}$ and subjecting it to quantum evolution that respects the structure of the subsystem code; for example, we can evolve the state using a generic local Hamiltonian $H$ constructed from the generators \eqref{eq:SymmetricOpsGen} of $G$. Because any operator in $G$ commutes with $\mathcal X$ and $\mathcal Z$, the quantum dynamics of the initial state preserves the state of the logical qubit. 
In this way the encoded qubit is protected by a combination of locality and symmetry; this is reminiscent of Kitaev's Majorana wire code~\cite{Kitaev01}, where a nonlocal but few body term that acts simultaneously on both ends of an open wire can enact a logical operator, but a combination of fermion parity symmetry and locality protects the encoded information. 

This reasoning is in fact quite general, and applies equally well to any locally-generated quantum dynamics that respects the Hilbert-space decomposition \eqref{eq:psi}. For example, one could consider dynamics of an initial state under a random unitary circuit $\prod_\mu e^{-i\theta_\mu\mathcal O_\mu}$, where $\mathcal O_\mu$ is any local operator in $G$ (or a linear combination thereof). One can even consider unitary-projective (or open-system) dynamics, so long as the projective measurements (or system-bath couplings) are also local operators drawn from $G$.  
The only constraints on the dynamics are (i) spatial locality and (ii) the topological symmetries \eqref{eq:TopoSymm}. This is similar to codes defined by symmetry-protected commuting projector Hamiltonians that protect against spatially local symmetric errors.

We formulate more precisely and elaborate further on the connections between subsystem codes and dynamics in Sec.~\ref{sec:SubsystemCodes}.

\subsection{Disentangling unitary}
\label{sec: Disentangling unitary}
\begin{figure}[t]
\includegraphics[width=1.00\columnwidth]{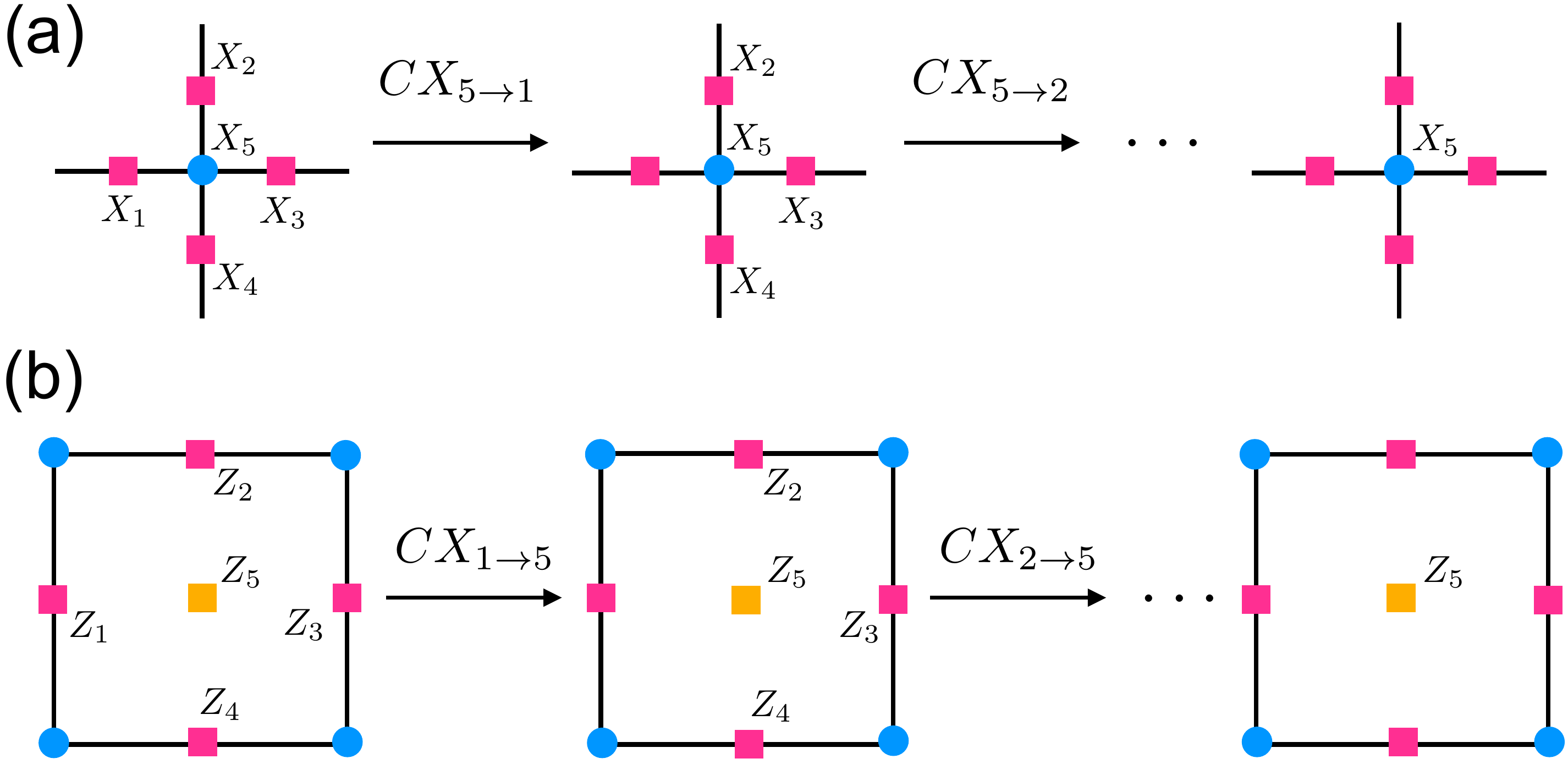}
\caption{
{\it Disentangling circuit.} 
        (a) The action of the disentangling circuit 
        $U_{v} = \prod_{l \in v} CX_{v \rightarrow \ell}$ on the qubits surrounding a vertex $v$. 
        (b) The action of the disentangling circuit $U_{p} = \prod_{l \in p} CX_{v \rightarrow p}$ on the qubits surrounding a plaquette $p$. 
        After application of the total circuit $U = \prod_{p}U_p \prod_{v}U_v$ we are left with matter qubits that are no longer dynamical. Thus, the ``gauge'' qubits are the only relevant degrees of freedom in the system. 
}
\label{fig:disentangle}
\end{figure}
In the next subsection, we exemplify the above discussion with numerical results on a concrete model Hamiltonian. To facilitate this study, it is useful to first apply a unitary disentangling circuit that effectively allows us to ``integrate out" the matter qubits, greatly reducing the number of degrees of freedom to be simulated. The circuit is a product of CNOT gates $CX_{i\to j}$, which act on Pauli operators as
\begin{align}
\begin{split}
    CX_{i\to j}\, X_i\, CX_{i\to j}&= X_i X_j\\
    CX_{i\to j}\, X_j\, CX_{i\to j}&= X_j\\
    CX_{i\to j}\, Z_j\, CX_{i\to j}&= Z_i Z_j\\
    CX_{i\to j}\, Z_i\, CX_{i\to j}&= Z_i.
\end{split}
\end{align}
For each vertex $v$, we apply the local unitary [see Fig.~\ref{fig:disentangle}(a)]
\begin{align}
    U_v = \prod_{\ell\in \ell_v}CX_{v\to\ell},
\end{align}
which acts on the gauge generators \eqref{eq:GaugeGenerators} as
\begin{align}
\label{eq:UvAction}
\begin{split}
    &U_v\, A_v\, U^\dagger_v = X_v \\
    &U_v\, B_p\, U^\dagger_v = B_p.
\end{split}
\end{align}
For each plaquette $p$, we apply the local unitary [see Fig.~\ref{fig:disentangle}(b)]
\begin{align}
    U_p = \prod_{\ell\in \ell_p}CX_{\ell\to p},
\end{align}
which acts on the gauge generators \eqref{eq:GaugeGenerators} as
\begin{align}
\label{eq:UpAction}
\begin{split}
    &U_p\, A_v\, U^\dagger_p = A_v \\
    &U_p\, B_p\, U^\dagger_p = Z_p.
\end{split}  
\end{align}
The total disentangling circuit,
\begin{align}
    U = \prod_p U_p\, \prod_v U_v,
\end{align}
thus maps the gauge-invariant Hilbert space $\mathcal H$ of Eq.~\eqref{eq:HilbertSpace}
to
\begin{align}
\label{eq:THilbertSpace}
     \tilde{\mathcal{H}}=\{\ket{\psi}\mid X_v\ket{\psi}=Z_p\ket{\psi}=\ket{\psi}\}.
\end{align}
In the transformed Hilbert space, the matter qubits are no longer dynamical and we can make the replacements $X_v,Z_p\mapsto+1$ 
$\forall\, v,p$.

We now consider the action of the disentangling circuit on the operators defining the subsystem code. We first consider the generators~\eqref{eq:SymmetricOpsGen} of $G$, the set of gauge-invariant local operators respecting topological symmetry~\eqref{eq:TopoSymm}. The transformed generators $X_v$ and $Z_p$ can be obtained by inverting Eqs.~\eqref{eq:UvAction} and \eqref{eq:UpAction}:
\begin{subequations}
\label{eq:XvZpAction}
\begin{align}
\begin{split}
    U\, X_v\, U^\dagger &= A_v = X_v\, \tilde A_v\equiv \tilde A_v\\
    U\, Z_p\, U^\dagger &= B_p = Z_p\, \tilde B_p\equiv \tilde B_p,
\end{split}
\end{align}
where in the final equality on both lines we have replaced the matter-qubit operators by their eigenvalues in the space $\tilde{\mathcal H}$. The operators
\begin{align}
    \tilde A_v=\prod_{\ell\in v}X_\ell,\indent \tilde B_p=\prod_{\ell\in p}Z_\ell
\end{align}
\end{subequations}
are the usual toric-code stabilizers. The remaining generators transform as
\begin{align}
\label{eq:XXX/ZZZAction}
\begin{split}
     U\, X_{p_1}X_{\ell(p_1,p_2)}X_{p_2}\, U^\dagger&=X_{\ell(p_1,p_2)}\\
     U\, Z_{v_1}Z_{\ell(v_1,v_2)}Z_{v_2}\, U^\dagger&=Z_{\ell(v_1,v_2)}.
\end{split}
\end{align}
Note that in the above expression, the links $\ell(p_1,p_2)$ and $\ell(v_1,v_2)$ must lie in the bulk of the lattice, since they are shared by two plaquettes and vertices, respectively. The generators~\eqref{eq:NonSymmetricOps} of $\overline{G}$, the set of gauge-invariant symmetry-violating operators, transform as
\begin{align}
\label{eq:XX/ZZAction}
\begin{split}
    U\, X_{\ell}X_{p(\ell)}\, U^\dagger &=X_\ell\\
    U\, Z_{\ell}Z_{v(\ell)}\, U^\dagger &=Z_\ell,    
\end{split}
\end{align}
where the link $\ell$ lies on a smooth boundary on the top line and a rough boundary on the bottom line.

Finally, we consider the action of the disentangling circuit on the topological symmetry generators \eqref{eq:TopoSymm} and on the logical operators \eqref{eq:BareLogicals}. From Eqs.~\eqref{eq:XvZpAction}, we find that the topological symmetry generators become
\begin{align}
\label{eq:TTopoSym}
\begin{split}
    \tilde S_X&=U\, S_X\, U^\dagger = \prod_{\ell\in\mathrm{rough}}X_\ell\\
    \tilde S_Z&=U\, S_Z\, U^\dagger = \prod_{\ell\in\mathrm{smooth}}Z_\ell,
\end{split}
\end{align}
where $\ell\in\mathrm{rough (smooth)}$ denotes the set of links belonging to the rough (smooth) boundaries. This mapping arises because all links belong to two toric-code stabilizers, except at the boundaries. Meanwhile, inverting Eqs.~\eqref{eq:XXX/ZZZAction} and \eqref{eq:XX/ZZAction} shows that the logical operators \eqref{eq:BareLogicals} are unchanged:
\begin{align}
\begin{split}
\label{eq:TLogicals}
    \tilde{\mathcal X} &= U\, \mathcal X\, U^\dagger=\mathcal X\\
    \tilde{\mathcal Z} &= U\, \mathcal Z\, U^\dagger=\mathcal Z\,.
\end{split}
\end{align}
Note that the logical operators are supported on only one rough or smooth boundary, while the topological symmetry operators are supported on \textit{both} rough or smooth boundaries. 
\begin{figure*}
            \includegraphics[width=1.00\textwidth]{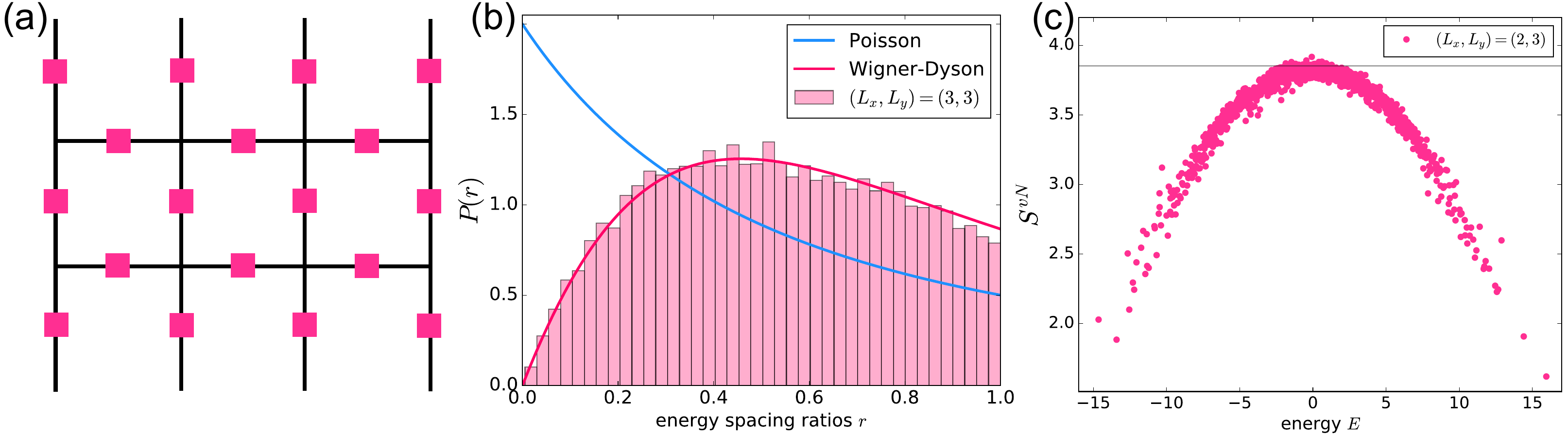}
\caption{
{\it Model with ``matter'' qubits integrated out.} 
         (a) A lattice of size $(L_x,L_y) =(3,3)$ with only the 18 ``gauge'' qubits is depicted. ``Matter'' qubits are suppressed after application of the disentangling circuit of Sec.~\ref{sec: Disentangling unitary}. 
         (b) Histogram $P(r)$ of energy spacing ratios $r$ with overlaid curves showing the analytical forms of the Wigner-Dyson GOE (magenta) and Poisson distributions (blue).
         The average value $\bar r$ of $P(r)$ is calculated to be $\approx 0.529$, while the analytical value is $\bar{r}_{\rm GOE}\approx 0.536$. 
        (c) The von-Neumann entropy $S^{vN}$ for all $8,192$  
        states in the spectrum for a lattice size of $(L_x,L_y) = (2,3)$. 
        The maximum value of $S^{vN}$ is close to the Page value $S_{\rm Page}\approx 3.85$ (see equation \eqref{page}), indicating that eigenstates near the middle of the spectrum have no special structure lending evidence to the fact that the Hamiltonian \eqref{eq:ModelHZ2} is generic. 
}
\label{fig:histo_entanglement}
\end{figure*}
\subsection{Concrete model Hamiltonian}
\label{sec:Z2Model}
We now define a concrete model Hamiltonian, constructed using the subsystem-code generators~\eqref{eq:SymmetricOpsGen}, that features a single encoded logical qubit as long as the topological symmetry is enforced. The Hamiltonian, defined in the Hilbert space $\tilde{\mathcal H}$ with matter qubits integrated out (see Sec.~\ref{sec: Disentangling unitary} and Fig.~\ref{fig:disentangle}), is given by
\begin{align}
\label{eq:ModelHZ2}
    H = \sum_v h_v\, \tilde A_v + \sum_p h_p\, \tilde B_p +\!\!\!\!\! \sum_{\ell\notin\mathrm{smooth}} \!\!\!\! J_{x,\ell}\, X_\ell+\!\!\!\!\sum_{\ell\notin\mathrm{rough}}\!\!\! J_{z,\ell}\, Z_\ell.
\end{align}
This Hamiltonian is reminiscent of that of the surface code (i.e., the toric code with boundary conditions as depicted in Fig.~\ref{fig:lattice}) in the presence of magnetic fields in the $x$- and $z$-directions. However, the model is constrained such that no $z$-fields appear on the rough boundaries, and no $x$-fields appear on the smooth boundaries. (We note, however, that one could include terms involving products of an even number of $X$- or $Z$-operators on the smooth or rough boundaries, respectively, without adversely affecting the encoded qubit.) These constraints ensure that the Hamiltonian \eqref{eq:ModelHZ2} obeys [see Eqs.~\eqref{eq:TTopoSym} and \eqref{eq:TLogicals}]
\begin{align}
    [H,\tilde S_X]=[H,\tilde S_Z]=[H,\tilde{\mathcal X}]=[H,\tilde{\mathcal Z}]=0.
\end{align}
In other words, irrespective of the values of the parameters $h_v,h_p,J_{x,\ell},J_{z,\ell}$---they can be arbitrarily large or small, and even vary in space---the model \eqref{eq:TTopoSym} respects the Hilbert-space decomposition \eqref{eq:psi} and therefore harbors an encoded logical qubit addressed by the logical operators $\tilde{\mathcal X}$ and $\tilde{\mathcal Z}$. 

It is important to note that the logical qubit preserved by the Hamiltonian \eqref{eq:ModelHZ2} is not the same as the one that appears in the surface/toric code. In the latter case, the encoded qubit resides in the ground state manifold and is only stable to small external fields. Here, instead, the logical operators that encode the qubit reside on the open boundaries of the lattice, and so are unaffected by even arbitrarily strong fields in the bulk. This gives rise to a twofold degeneracy of \textit{every} energy eigenstate, rather than just the ground state. Indeed, if the model is defined with periodic rather than open boundary conditions, the encoded qubit disappears. In this respect, the encoded qubit in this model more closely resembles the qubits encoded by strong zero modes in 1D systems~\cite{Fendley05,Fendley16,Else17a,Kemp17}.

    Despite the fact that it respects the Hilbert-space decomposition $\eqref{eq:psi}$ and therefore encodes a logical qubit, the Hamiltonian \eqref{eq:ModelHZ2} is quite generic by standard measures. To emphasize this, we present numerical exact diagonalization (ED) results for the model \eqref{eq:ModelHZ2} defined on the lattice depicted in Fig.~\ref{fig:lattice}(a). 
    After integrating out all matter qubits, we are left with a model 
    of interacting gauge qubits as shown in Fig.~\ref{fig:histo_entanglement}(a). 
    We label system sizes by the number of plaquettes in the horizontal and vertical directions, which we call $L_x$ and $L_y$, respectively. We use model parameters $J_{x,\ell}=J_{z,\ell}=1$, and random $h_v$ and $h_p$ drawn uniformly from the interval $[1-0.2,1+0.2]$. All data shown are for a single disorder realization.

In Fig.~\ref{fig:histo_entanglement}(b), we show that the energy level statistics of the model fits predictions from random matrix theory, as expected for nonintegrable Hamiltonians~\cite{D'Alessio16}. We plot the probability distribution $P(r)$ of the level spacing ratio~\cite{Oganesyan07,Pal10}
\begin{align}
 r_i = \frac{\text{min}(E_{i+1}-E_{i},E_{i+2}-E_{i+1})}{\text{max}(E_{i+1}-E_{i},E_{i+2}-E_{i+1})}, 
\end{align}
where $\{E_i\}$ is an ordered list of the eigenvalues of $H$. The data shown are for system size $(L_x,L_y)=(3,3)$ for energy levels in the topological symmetry sector $(\tilde S_X,\tilde S_Z)=(+1,+1)$, upon restricting to states with $\tilde{\mathcal Z}$ eigenvalue $+1$ ($32,768$ states in total). The resulting histogram is in excellent agreement with the analytical expression for $P(r)$ in the Gaussian Orthogonal Ensemble (GOE) of random matrices~\cite{Atas13}. For comparison, the average value $\bar r$ of $P(r)$ in Fig.~\ref{fig:histo_entanglement}(b) is $\approx 0.529$, while the analytical value is $\bar{r}_{\rm GOE}\approx 0.536$. Since integrable systems exhibit Poisson level statistics with $\bar{r}_{\rm Poisson}\approx0.386$, this is strong evidence that the Hamiltonian $\eqref{eq:ModelHZ2}$ is generic once all commuting conserved quantities are accounted for.
 
In Fig.~\ref{fig:histo_entanglement}(c), we plot the von Neumann entanglement entropy $S^{vN}=-\text{tr}(\rho_A\ln\rho_A)$,
where $\rho_A=\text{tr}_B(\ket{E}\bra{E})$ is the reduced density matrix of half of the system, for all eigenstates $\ket{E}$ in the $(\tilde S_X,\tilde S_Z,\tilde{\mathcal Z})=(+1,+1,+1)$ sector at system size $(L_x,L_y)=(2,3)$. The entanglement-vs.-energy curve displays a ``domed'' structure characteristic of models obeying the ETH. The maximum value of the curve is close to the ``Page value''~\cite{Page93}
\begin{align}\label{page}
    S_{\rm Page}=N_A\ln 2-\frac{2^{N_A}}{2^{N_B+1}}
\end{align}
characteristic of a random state with $N_A$ qubits in region $A$ and $N_B$ qubits in region $B$. For the bipartition used in Fig.~\ref{fig:histo_entanglement}(c), $N_A=7$ and $N_B=6$, so that $S_{\rm Page}\approx 3.85$. This indicates that eigenstates near the middle of the many-body spectrum have no special structure apart from the $O(1)$ number of commuting conserved quantities, further substantiating the fact that the Hamiltonian \eqref{eq:ModelHZ2} is generic.

Taken together, these numerical results substantiate the surprising fact that a commuting pair of global symmetries, $\tilde S_X$ and $\tilde S_Z$, is sufficient to encode a logical qubit in the Hilbert space of an otherwise generic model. They also strongly suggest that the system obeys the ETH once the commuting conserved quantities $(\tilde S_X,\tilde S_Z,\tilde{\mathcal Z})$ [or, equivalently, $(\tilde S_X,\tilde S_Z,\tilde{\mathcal X})$] are specified. It is therefore expected that a quantum quench in which the system is prepared in an arbitrary state of the form \eqref{eq:psi} (e.g., a $Z$-basis product state) will generically yield relaxation to a state that locally resembles a Gibbs ensemble with fixed values of these conserved quantities~\cite{D'Alessio16}. The same phenomenology applies to initial states where $\ket{\psi_{\rm L}}$ occupies a generic position on the logical Bloch sphere, i.e., where $\ket{\psi_{\rm L}}$ is the $+1$ eigenstate of a logical operator of the form $n_x\, \tilde{\mathcal X} + n_y\, \tilde{\mathcal Y}+n_z \tilde{\mathcal Z}$, where $\tilde{\mathcal Y}=i\tilde{\mathcal X}\tilde{\mathcal Z}$. Such initial states can be prepared using techniques developed for surface codes, see e.g.~Refs.~\cite{Brown17,Satzinger21}.


\subsection{Stability and Prethermalization} 
\label{sec:Prethermalization}


Having uncovered the structure of a symmetry-protected quantum memory within an otherwise quantum-chaotic many-body system, the question naturally arises of whether this memory can be stabilized against perturbations. Naively, it would seem that any perturbation to Eq.~\eqref{eq:ModelHZ2} with energy scale $g$ that violates the conservation of $\tilde S_{X/Z}$ (or equivalently $S_{X/Z}$ in the language of Eq.~\ref{eq:TopoSymm}) would lead to decoherence 
of the encoded qubit in a time $t_*\sim 1/g$. However, it turns out that the lifetime of the quantum memory can be extended far beyond this naive limit by leveraging the concept of prethermalization~\cite{Abanin17,Else17a}. Invoking the results of Ref.~\cite{Else20}, we argue that the lifetime can be bounded from below by a stretched exponential,
\begin{align}
\label{eq:t*m=2}
    t_*\gtrsim e^{C\sqrt{\nu/g}},
\end{align}
where $C$ is a constant and where $\nu\gg g$ is an energy scale, explained below, that suppresses the creation of excitations that could cause bit-flip or phase errors in the encoded qubit. The discussion in this Section follows Refs.~\cite{Else17a,Else20}, which consider a weakly perturbed surface code and find the same long-lived encoded qubit. The primary difference between our work and Refs.~\cite{Else17a,Else20} is that our starting point is the family of nonintegrable Hamiltonians defined in Sec.~\ref{sec:Z2Model}, which follows from the subsystem-code structure elucidated at the beginning of Sec.~\ref{sec:Z2code}.

In order to leverage prethermalization, we embed the $\mathbb Z_2\times\mathbb Z_2$ topological symmetry group generated by Eq.~\ref{eq:TopoSymm} into a U(1)$\times$U(1) symmetry group with generators
\begin{subequations}
\begin{align}
    N_X&=\sum_v X_v,\indent N_Z=\sum_p Z_p,
\end{align}
or equivalently, in the disentangled picture,
\begin{align}
    \tilde N_X=\sum_v \tilde A_v,\indent \tilde N_Z&=\sum_p \tilde B_p.
\end{align}
\end{subequations}
These operators count the number of vertex and plaquette excitations of the surface code; the original topological symmetry generators $S_{X,Z}$ measure the parity of $N_{X,Z}$, respectively. It is possible to write down an analog of Eq.~\eqref{eq:ModelHZ2} that respects this enlarged topological symmetry,
\begin{widetext}
\begin{align}
\label{eq:ModelHU1}
    H_0 = \sum_v h_v\, \tilde A_v + \sum_p h_p\, \tilde B_p +\!\!\!\! \sum_{\ell\notin\mathrm{smooth}} \!\!\!\! J_{x,\ell}\,\left(\frac{1-\tilde B_{p_1(\ell)}\tilde B_{p_2(\ell)}}{2}\right) X_\ell+\!\!\!\!\sum_{\ell\notin\mathrm{rough}}\!\!\! J_{z,\ell}\, \left(\frac{1-\tilde A_{v_1(\ell)}\tilde A_{v_2(\ell)}}{2}\right)Z_\ell,
\end{align}
\end{widetext}
where $p_{1,2}(\ell)$ are the two plaquettes that share link $\ell$ and $v_{1,2}(\ell)$ are the two vertices that share link $\ell$. This model arises as a projection of Eq.~\eqref{eq:ModelHZ2} into a sector with a fixed number of vertex and plaquette excitations. We have checked numerically using the methodology of Sec.~\ref{sec:Z2Model} that the model \eqref{eq:ModelHU1} remains nonintegrable while retaining the same encoded logical qubit. For simplicity, we henceforth restrict to the case of uniform parameters $h_v=h_p=0$ and $J_{x,\ell}=J_{z,\ell}=J$.
\begin{figure*}[t!]
\includegraphics[width=1.00\textwidth]{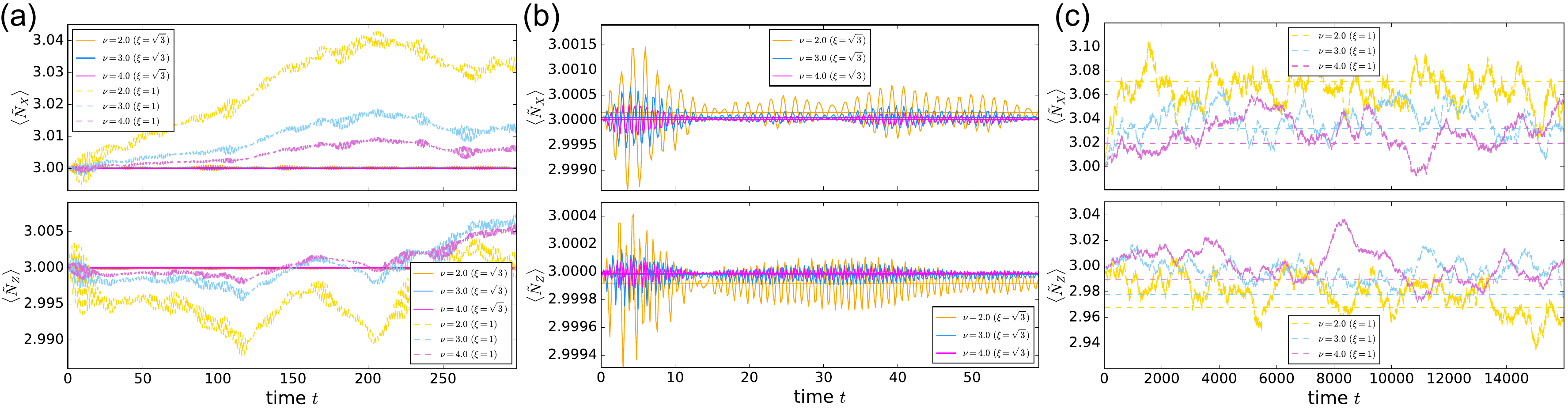}
\caption{
{\it Dynamics of the expectation values of the vertex and plaquette excitation operators $\braket{\tilde N_{X}}$ [top panel] and $\braket{\tilde N_{Z}}$ [bottom panel].}  
(a) The dynamics of $\braket{\tilde N_{X}}$ [top panel] and $\braket{\tilde N_{Z}}$ [bottom panel] from initial product states with $(\braket{\tilde N_{X}},\braket{\tilde N_{Z}})=(3,0)$ and $(0,3)$, respectively, are shown from time $t = 0$ until $t = 300$. Parameters used in the plots are $(L_x,L_y)=(2,3)$, $J=0.1$, and $g=0.15$. Yellow, blue, and magenta curves use parameters $\nu=(2.0,\,3.0,\,4.0)$, respectively. Dashed curves use $\xi=1$, while solid curves belong to  $\xi=\sqrt{3}$. Solid (dashed) horizontal lines indicate the diagonal ensemble averages $\braket{\tilde N_{X(Z)}}_{\rm DE}$ for 
$\xi = \sqrt{3}$ $(\xi = 1.0)$. 
(b) We plot the data for the case $\xi = \sqrt{3}$ shown in (a) again for a smaller range of time $t$, i.e. $t = 0, \ldots, 60$. As easily seen now, the expectation values $\braket{\tilde N_{X}}$  and $\braket{\tilde N_{Z}}$ display oscillating behavior around the diagonal ensemble averages 
$\braket{\tilde N_{X,Z}}_{\rm DE}$ with 
$\braket{\tilde N_{X}}_{\rm DE} = (3.00015,\,3.00005,\,3.00002)$ and 
$\braket{\tilde N_{Z}}_{\rm DE} = (2.99992,\,2.99998,\,2.99999)$ 
for $\nu = (2.0,\,3.0,\,4.0)$, respectively. 
On the qualitative side we note that 
with increasing factor $\nu$, the 
oscillations become stronger. 
(c) We spotlight the behavior of $\braket{\tilde N_{X}}$  and $\braket{\tilde N_{Z}}$ for the case of $\xi = 1.0$. Contrarily to the case $\xi = \sqrt{3}$ detailed in (b), we now observe distinctively different behavior. The computed expectation values 
$\braket{\tilde N_{X(Z)}}$ do not show any oscillations around their diagonal ensemble averages. Instead we see a strong discrepancy between $\braket{\tilde N_{X(Z)}}$ and 
$\braket{\tilde N_{X(Z)}}_{\rm DE}$. 
To give the eye a guide line we again include the diagonal ensemble averages $\braket{\tilde N_{X(Z)}}_{\rm DE}$ as dashed lines. 
Their exact values are 
$\braket{\tilde N_{X}}_{\rm DE} = (3.07146,\,3.03196,\,3.01963)$ and 
$\braket{\tilde N_{Z}}_{\rm DE} = (2.96784,\,2.97793,\,2.98982)$ for 
$\nu = (2.0,\,3.0,\,4.0)$, respectively. 
Thus, the presented numerical data shows that emergent 
U(1)$\times$U(1) symmetry persists to long times $t$ for 
irrational $\xi$, but not for integer $\xi$,  consistent with the predictions of the theory of prethermalization.
}
\label{fig:resonances}
\end{figure*}

Next, we consider the effect of symmetry-violating perturbations on the symmetric model \eqref{eq:ModelHU1}. These perturbations are of the form
\begin{align}
\label{eq:U1Pert}
    V=g_x\sum_{\ell\in\text{smooth}}X_\ell+g_z\sum_{\ell\in\text{rough}}Z_\ell.
\end{align}
Physically, these perturbations create individual vertex and plaquette excitations at the appropriate boundaries of the system. We henceforth simplify these perturbations by setting $g_x=g_z=g$. Note that the perturbations \eqref{eq:U1Pert} reside on the boundaries of the lattice and break the $\mathbb Z_2\times\mathbb Z_2$ subgroup of the U(1)$\times$U(1) symmetry. Because the generators of the $\mathbb Z_2\times\mathbb Z_2$ subgroup reside on the boundaries, any local perturbation added to the bulk will preserve this subgroup, even if it breaks the U(1)$\times$U(1) symmetry. Since such local U(1)$\times$U(1)-breaking but $\mathbb Z_2\times\mathbb Z_2$-preserving bulk terms do not affect the logical qubit (see also Ref.~\cite{Else17a}), we opt to focus instead on the boundary perturbations~\eqref{eq:U1Pert}, which are detrimental to the encoded qubit.

Prethermalization aims to suppress the effect of symmetry-violating perturbations [e.g.,~Eq.~\eqref{eq:U1Pert}] by inducing a separation of energy scales~\cite{Abanin17}. To achieve this, we add two large ``external fields" that couple to the two U(1)$\times$U(1) generators; i.e., we consider a Hamiltonian of the form
\begin{align}
\label{eq:FullModelHU1}
    H =\nu_x \tilde N_X+\nu_z \tilde N_Z+ H_0+V
\end{align}
in the limit $\nu_{x,z}\gg g$. This imposes an energetic penalty on violations of the U(1)$\times$U(1) symmetry. To simplify the analysis, we will write 
\begin{align}
\label{eq:xi-def}
\nu_z = \xi\, \nu_x \equiv \xi\, \nu,
\end{align}
where $\xi$ is an order-one number to be specified later. Deviations from $\tilde N_{X,Z}$ conservation can then be captured perturbatively in $g/\nu$ (in the absence of resonances, discussed below) by an effective Hamiltonian that is U(1)$\times$U(1)-symmetric at each order. The validity of this type of perturbation theory has been rigorously addressed for the case of U(1) symmetry~\cite{Abanin17} and for the more general case of $K$ commuting U(1) symmetries~\cite{Else20}. In this case, it can be shown that the perturbation theory is valid up to an order
\begin{subequations}
\label{eq:Timescale-General-m}
\begin{align}
    n_{*}\sim\left(\frac{\nu}{g}\right)^{1/K},
\end{align}
corresponding to a timescale
\begin{align}
    t_{*}\gtrsim e^{C(\nu/g)^{1/K}}
\end{align}
\end{subequations}
on which the $K$ U(1) charges are approximately conserved.
For a single U(1) symmetry, this corresponds to an exponentially long-lived approximate conservation law. The Hamiltonian~\eqref{eq:FullModelHU1} falls into the special case $K=2$, where the stretched exponential timescale \eqref{eq:t*m=2} is recovered.


The results of Ref.~\cite{Else20} can only be applied when the external fields coupling to the U(1) charges are \textit{irrational} multiples of one another. Otherwise, the U(1)$^{\times K}$-symmetric perturbation theory leading to Eqs.~\eqref{eq:Timescale-General-m} is spoiled by resonant processes that change the U(1) charges in such a way that the net energy change due to the external fields vanishes. In the model \eqref{eq:FullModelHU1}, this occurs when $\nu_{x}$ and $\nu_{z}$ are rational multiples of each other, i.e., when $\xi=p/q$ for some relatively prime $p,q\in\mathbb Z$ in Eq.~\eqref{eq:xi-def}. In this case, one can have a resonant process in which, e.g., $\tilde N_Z\to \tilde N_Z+q$ and $\tilde N_X\to \tilde N_X-p$, such that the external-field term $\nu(\tilde N_X+\xi \tilde N_Z)$ is invariant. If allowed, such ``resonances" would lead to decay of the encoded qubit on a timescale independent of $\nu$, since the qubit encoded in $H_0$ is protected by the U(1)$\times$U(1) symmetry rather than the U(1) subgroup generated by $q\tilde N_X+p\tilde N_Z$. Setting $\xi$ to an irrational number essentially suppresses these processes, endowing the dynamics under Eq.~\eqref{eq:FullModelHU1} for $g\ll \nu$ with a long-lived approximate U(1)$\times$U(1) symmetry.


Numerical results probing the emergent U(1)$\times$U(1) symmetry and the effect of resonances are shown in Fig.~\ref{fig:resonances}. Using ED, we calculate the dynamics from representative initial product states of $\braket{\tilde N_{X}}$ and $\braket{\tilde N_{Z}}$ for several values of $\nu$ for both $\xi=1$ and $\xi=\sqrt{3}$. For $\xi=1$, $\braket{\tilde N_{X}}$ and $\braket{\tilde N_{Z}}$ immediately begin to exhibit strong discrepancies in comparison with  their initial values, indicative of the resonant processes discussed above. For $\xi=\sqrt{3}$, however, $\braket{\tilde N_{X}}$ and $\braket{\tilde N_{Z}}$ do not strongly diverge indefinitely from their initial values. Contrarily, they begin to oscillate around saturation values close to their initial values. These saturation values are well-described by the ``diagonal ensemble''~\cite{Kollar08,Rigol08,D'Alessio16}, defined for an arbitrary operator $\mathcal O$ in the eigenbasis $\ket{E}$ of $H$ as
\begin{align}
    \braket{O}_{\rm DE} = \sum_E |\braket{\psi_0|E}|^2\braket{E|O|E},
\end{align}
where $\ket{\psi_0}$ is the initial state. Physically, $\braket{O}_{\rm DE}$ is the infinite-time average of $\braket{O(t)}$; the fact that $\braket{\tilde N_{X,Z}}_{\rm DE}$ are so close to the values of $\braket{\tilde N_{X,Z}}$ at time $t=0$, and in fact become closer as $\nu$ increases, is indicative of the emergent U(1)$\times$U(1) conservation law in the limit ${\nu/g\to\infty}$. In particular, the fact that $\braket{\tilde N_{X,Z}}$ stabilize around these values after an initial transient suggests that the dynamics of the system can be well approximated by one that conserves both $\tilde N_{X}$ and $\tilde N_{Z}$.

\begin{figure}[t]
\includegraphics[width=1.00\columnwidth]{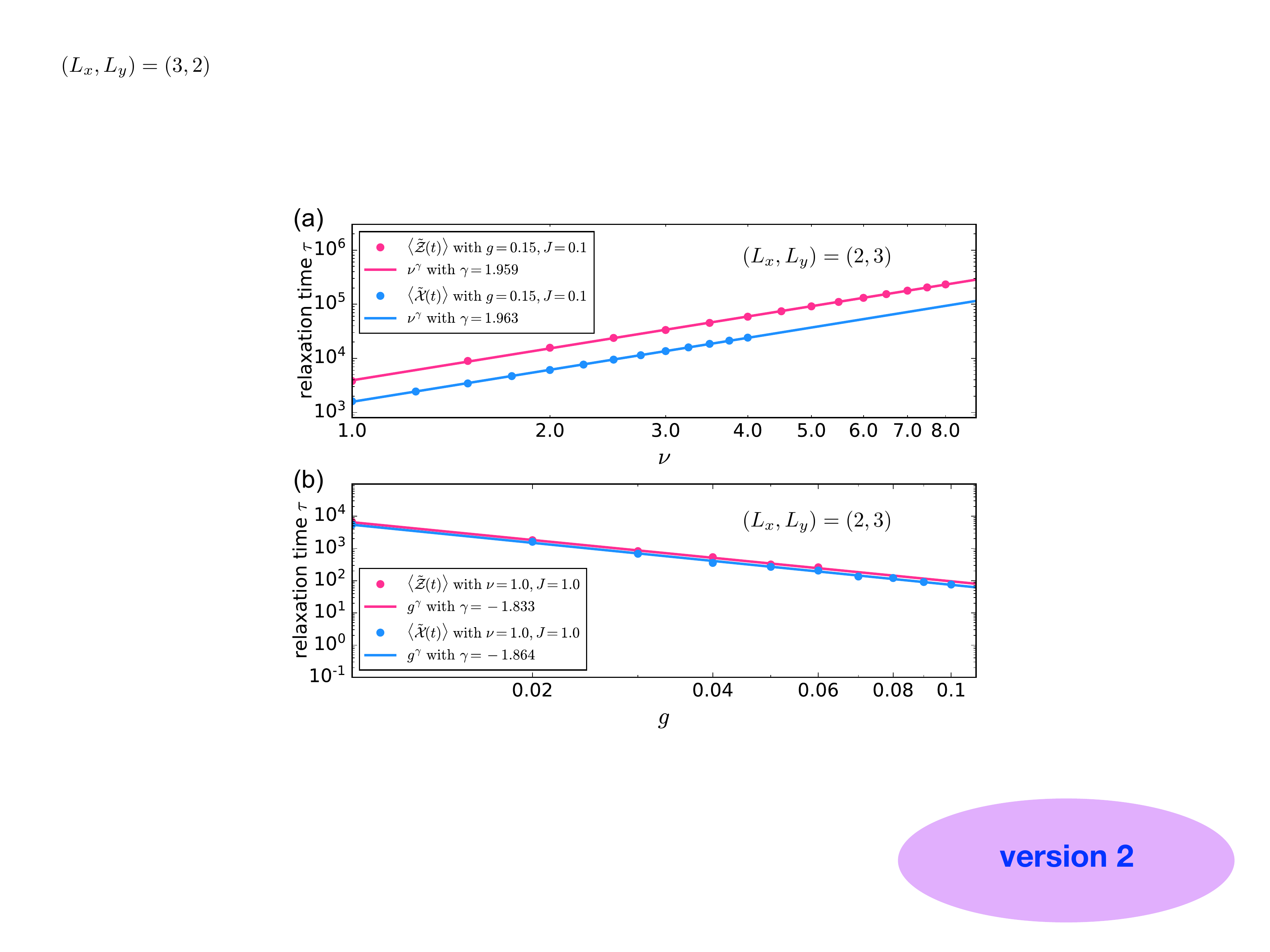}
\caption{
{\it Scaling behavior of relaxation times $\tau_{Z}$ and $\tau_{X}$.} 
        Scaling of the lifetimes $\tau_{Z}$ (magenta data points) and $\tau_{X}$ (blue data points) with the coupling paramters $\nu$  [top panel (a)] and $g$ [bottom panel (b)]   extracted from scrutinizing the quench dynamics of $\braket{\tilde{\mathcal X}(t)}$ and $\braket{\tilde{\mathcal Z}(t)}$ 
        starting from initial product states, i.e. 
        eigenstates of the respective logical operator. The lifetime $\tau_{Z(X)}$ is then defined to be the time where 
        $\braket{\tilde{\mathcal X}(t)}=\braket{\tilde{\mathcal X}}_{\rm DE}$ [$\braket{\tilde{\mathcal Z}(t)}=\braket{\tilde{\mathcal Z}}_{\rm DE}$] is true for the first time. 
        The coupling parameter $\nu$ governs the strength of the operators $\tilde N_X$ and $\tilde N_Z$ that  count  the  number  of  vertex  and  plaquette excitations of the surface code whereas the parameter $g$ is associated with the  symmetry-violating  perturbation $V$ (see \eqref{eq:U1Pert}).  
        System size is again $(L_{x},L_{y})=(2,3)$ and initial states are the same as those used in Fig.~\ref{fig:resonances}, and we use $\xi=\sqrt{3}$ to avoid resonances. Data points are plotted on a log-log scale and fits to power-law scaling are shown as solid lines.
        The powers extracted from the fits are $1.959$ ($1.963$) for $\braket{\tilde{\mathcal Z}(t)}$ [$\braket{\tilde{\mathcal X}(t)}$], and
        $-1.833$ ($-1.864$) for $\braket{\tilde{\mathcal Z}(t)}$ [$\braket{\tilde{\mathcal X}(t)}$].
}
\label{fig:tau}
\end{figure}


To probe the lifetime of the encoded qubit in the presence of perturbations, we simulate the quench dynamics of $\braket{\tilde{\mathcal X}(t)}$ and $\braket{\tilde{\mathcal Z}(t)}$ starting from initial product states that are eigenstates of one of the logical operators. We calculate the lifetime $\tau_X$ ($\tau_Z$) as the first time at which $\braket{\tilde{\mathcal X}(t)}=\braket{\tilde{\mathcal X}}_{\rm DE}$ [$\braket{\tilde{\mathcal Z}(t)}=\braket{\tilde{\mathcal Z}}_{\rm DE}$]. We then repeat the simulation for a sequence of $g$ and $\nu$ values to extract the dependence of $\tau_{X,Z}$ on these parameters. 
The data, shown in Fig.~\ref{fig:tau}, show a clear \textit{power-law} scaling in $\nu/g$, rather than the expected stretched-exponential scaling of Eq.~\eqref{eq:t*m=2}.

This deviation from the expected scaling can be understood straightforwardly as a finite-size effect. In the presence of (approximate) $\tilde N_{X,Z}$ conservation, only one type of process can ``flip'' the state of the logical qubit: an excitation is created at one of the boundaries, propagates through the bulk of the system, and is annihilated at the other boundary of the same (rough/smooth) type. If the excitation propagates between rough (smooth) boundaries, $\braket{\tilde{\mathcal X}}$ ($\braket{\tilde{\mathcal Z}}$) changes sign. This process is second-order in $g$, the energy scale for quasiparticle creation at the boundaries, and subextensive in the kinetic energy scale $J$, which propagates the excitation between boundaries. The relaxation time associated with this process for a system of size $(L_x,L_y)$ is then
\begin{eqnarray}
\label{eq:finite-size lifetime}
    \tau_{X} \sim \frac{\nu}{g^{2}}\left(\frac{\nu}{J}\right)^{L_{y}-2},
    \indent
    \tau_{Z} \sim \frac{\nu}{g^{2}}\left(\frac{\nu}{J}\right)^{L_{x}-1}.
\end{eqnarray}
In this formula, we omit constant factors and have included appropriate powers of $J$ for dimensional purposes. For sufficiently large $\nu$, these lifetimes diverge exponentially in the thermodynamic limit $L_x,L_y\to\infty$.
This recovers the familiar fact~\cite{Kitaev2003,Bravyi98,Fowler12,Fendley16,Else17a,Kemp17} that the encoded qubit is only stable up to exponentially small finite-size corrections (or, equivalently, exponentially long times at finite size). At the small system size $(L_x,L_y)=(2,3)$ considered here (due to runtime constraints, as each relaxation-time calculation involves evolving the system over times of up to $\sim10^4 J^{-1}$), these timescales both scale as $\nu^2/g^2$, consistent with the timescales observed in our numerics. Nevertheless, since the results of Fig.~\ref{fig:resonances} suggest that the approximate U(1)$\times$U(1) symmetry is operative, we expect that the regime in which $\tau\gtrsim t_*$ [Eq.~\eqref{eq:t*m=2}] holds should be accessible at larger system sizes, even if numerical simulation of the system at such sizes becomes infeasible. 

In summary, while our numerical results are hindered by finite-size effects, they clearly reveal the presence of a long-lived emergent U(1)$\times$U(1) symmetry despite the presence of symmetry-violating perturbations. These results are consistent with the existence of an encoded qubit with a lifetime parametrically long in the inverse of the perturbation energy scale, but larger-scale numerics would be required to validate the prediction of a stretched-exponential lifetime in this model.



\section{Subsystem Codes and Dynamics}
\label{sec:SubsystemCodes}


In this section we formalize our notion of quantum memory in a dynamical system by giving a definition in terms of noiseless subsystems. 
We then explain how subsystem codes provide plentiful nontrivial examples of such dynamical systems and uncover the appropriate local error detection structure required for these codes to give rise to prethermal quantum memories when the discrete stabilizer symmetry is enlarged to a U(1)$^{\times K}$ symmetry~\cite{Else20}. 


\subsection{Noiseless subsystems}
\label{sec:Noiseless subsystems}

A general, potentially open-system, quantum dynamical evolution on a Hilbert space $\mathcal{H}$ is described by a completely positive trace preserving (CPTP) map~\cite{Nielsen2010} $\mathcal{E}_t$ on the space of operators $\mathcal{B}(\mathcal{H})$ that is a function of a discrete or continuous time parameter $t$. 
Here, we are interested in dynamical evolutions that retain a quantum memory of initial conditions. 
A very general formulation of the quantum memory condition is obtained by requiring the dynamical evolution to be correctable with respect to a chosen quantum code.  
More precisely, quantum memory of a subsystem $\mathcal{H}_L$ within the full Hilbert space $\mathcal{H}= (\mathcal{H}_L \otimes \mathcal{H}_J) \oplus \mathcal{H}_\perp$---where $L,J,$ stand for logical and junk subsystems, respectively, and where $\perp$ denotes the orthogonal complement subspace---is retained if the dynamics under $\mathcal{E}_t$ is correctable.
That is, there exists a recovery map $\mathcal{R}_t$ such that~\cite{Kribs2005,Kribs2006}
\begin{align}
    \text{Tr}_J \circ P \circ \mathcal{R}_t \circ \mathcal{E}_t \, (\rho) =  \text{Tr}_J (\rho) \, ,
\end{align}
for $\rho$ a density matrix supported on $\mathcal{H}_L \otimes \mathcal{H}_J$ and $P$ a projector onto $\mathcal{H}_L \otimes \mathcal{H}_J$. 
This definition says that $\mathcal{H}_L$ forms a noiseless subsystem~\cite{Kribs2005,Kribs2006} for $\mathcal{R}_t\circ\mathcal{E}_t$. In particular, when $\mathcal{R}_t=\mathbbm{1}$ this reduces to $\mathcal{H}_L$ forming a noiseless subsystem for the dynamics $\mathcal{E}_t$. 
For the interesting subclass of closed-system unitary dynamics, the former condition is too weak to be nontrivial, since all unitary evolutions can in principle be undone. In this setting, we therefore focus on the latter definition, where the recovery map is taken to be $\mathcal{R}_t=\mathbbm{1}$ and $\mathcal{H}_L$ is a noiseless subsystem for $\mathcal{E}_t$, i.e. some initial quantum information remains static under the evolution. 

While the above definition applies to perfect quantum memory for arbitrarily long times, it is interesting to loosen this to an approximate quantum memory with high probability for a sufficiently long time, as we consider in a certain setting below. 
It is also easy to come up with trivial examples of dynamics that retain a quantum memory, such as a single qubit with trivial evolution. We are interested in nontrivial examples where every qubit participates in the dynamics to form a quantum memory via a collective quantum many-body effect. We further want this memory to be stabilizable against generic uniform local perturbations that remain sufficiently weak.

\subsection{Stabilizer subsystem codes} 
\label{sec:Stabilizer subsystem codes}

To construct specific examples of nontrivial noiseless subsystems, we focus our attention on stabilizer subsystem codes~\cite{Poulin2005,Bacon2005a} (see also Ref.~\cite{Knill00}). For the remainder of this section we restrict our attention to qubits and $\mathbb{Z}_2$ local degrees of freedom (i.e., qubits) for simplicity of presentation; the extension to $\mathbb{Z}_p$ for $p$ prime (i.e., qudits with prime dimension) is straightforward. Further extension to general $\mathbb{Z}_N$ is possible, however the theory of stabilizer subsystem codes for composite dimensions becomes more complicated. 

stabilizer subsystem codes are defined by a group of ``gauge'' operators $G\leq  P$ that form a subgroup of the Pauli group $P$. We remark that these ``gauge'' operators are not related to the gauge transformations of lattice gauge theory, but rather refer to operators acting on redundant ``gauge'' qubit degrees of freedom associated to $\mathcal{H}_J$. 
An auxiliary group of stabilizer operators $S$ is given by the center of the gauge group, $S=Z(G) < P$. This induces a decomposition of the Hilbert space 
\begin{align}
\label{eq:SubsystemCodeBlockDecomposition}
    \mathcal{H} = \bigoplus_{\{\lambda_i\}} (\mathcal{H}_L \otimes \mathcal{H}_J)_{\{\lambda_i\}}
\end{align}
into isomorphic sectors labelled by a complete set of stabilizer eigenvalues, specified here by the eigenvalues $(-1)^{\lambda_i}$ of a generating set of stabilizers $s_i$. Here $L$ denotes the ``logical subsystem" and $J$ denotes the ``junk subsystem" for reasons we now explain. 
The ``gauge'' operators act within each sector as $\mathbbm{1}_L\otimes g_J$. 
Operators in the commutant (i.e., centralizer) of the gauge subgroup $C(G)$ are called bare logical operators; their actions are only defined up to multiplication with elements of the stabilizer group. Therefore $L=C(G)/S$ describes the group of bare logical operators up to equivalence. These bare logical operators act within each sector as $\ell_L\otimes\mathbbm{1}_J$. 
As logical quantum information is only stored in the logical subsystem $\mathcal{H}_L$, a larger class of ``dressed logical" operators are relevant. These dressed logical operators implement the same set of transformations on $\mathcal{H}_L$ while also potentially disturbing $\mathcal{H}_J$ by an unimportant operator. The algebra of dressed logical operators is given by the commutant of the stabilizer subgroup, $C(S)$. Operators in $C(S)$ have a well defined logical action up to multiplication with ``gauge'' operators. Hence $C(S)/G$ is the group of dressed logical operators up to equivalence. Representative dressed logical operators act within each sector as $\ell_l \otimes g_J$. The code distance $d$ is given by the minimum Pauli weight over all representative dressed logical operators. The group $L=C(G)/S$ of bare logical operators is a projective representation of the group $\mathbb{Z}_2^{\times |L|}$, where $|L|$ is twice the number of encoded qubits. 
Similarly the group $C(S)/G$ of dressed logical operators, modulo complex phases, forms a representation of the same group $\mathbb{Z}_2^{\times |L|}$. 

The essential features of a stabilizer subsystem code are often packaged into the notation $[[n,k,r,d]]$ where $n$ denotes the number of physical qubits, $k$ the number of encoded logical qubits, $r$ the number of gauge qubits, and $d$ the code distance. In this notation, $[[n,k,0,d]]$ codes are conventional stabilizer codes. 
Quantum error correction for stabilizer subsystem codes is implemented by measuring the gauge generators (in a sequence that accounts for the fact they may not commute) and using this information to reconstruct the stabilizer eigenvalues. The measured stabilizer eigenvalues are then used to compute a correction operator that is applied to return the system to the code space such that the inferred error operator composed with the correction operator results in a trivial logical operation.

For concreteness we always fix a choice of generators for the gauge group, i.e., we write $G=\langle \{ g_i \} \rangle$, where the list of generators $g_i$ may be overcomplete. 
If the generators can be picked to consist of purely $X$-type and $Z$-type Pauli operators, the code is known as a CSS subsystem code~\cite{Calderbank1996,Steane1996}. 
Our focus is on systems with generators $g_i$ that respect the microscopic locality of an underlying lattice. 
Furthermore, we fix a choice of generators for the stabilizer group, $S=\langle \{ s_i \} \rangle$, which can involve a combination of local and nonlocal operators.
A \textit{topological subsystem code}~\cite{Bombin2009,Kargarian2010,Bombin10,Suchara11,Bravyi2013} is one for which the stabilizer generators can all be picked to be local with respect to an underlying lattice while the code distance $d$ is macroscopic (i.e., $d$ grows as the number of physical qubits increases). 

\subsection{Stabilizer subsystem code symmetry and dynamics}
\label{sec:SubsystemCodeDynamics}

In this subsection we discuss an interpretation of the stabilizer group of a subsystem code as an anomalous symmetry. We go on to describe the consequences that such a symmetry has on closed an open system dynamics whose generators are symmetric, of which the example in Sec.~\ref{sec:Z2code} is a special case. 

The stabilizer group of a nontrivial subsystem code forms a type of \textit{anomalous symmetry}. For our purposes here, an anomalous symmetry is defined by the property that no local symmetric gapped Hamiltonian can have a  unique ground state. 
More precisely, for nontrivial $\mathcal{H}_L$ of dimension $2^{|L|/2}$, any Hamiltonian that is given by a sum of gauge operators, i.e., 
\begin{align}
\label{eq:GaugeOpHam}
    H = \sum_{v} h_v \, ,
\end{align}
where the interactions $h_v$ are (linear combinations of) gauge operators, has at least a $2^{|L|/2}$-fold degeneracy of its energy levels as it commutes with the stabilizer and bare logical operators of the code. For the class of subsystem codes we consider, namely those with local generators that protect against local errors, any local Hamiltonian that commutes with the stabilizer symmetry must also commute with the bare logical operator algebra and hence must have degenerate energy levels. 
That is, any local Hamiltonian that respects the stabilizer symmetry group $S$, which is a linear representation of $\mathbb{Z}_2^{\times |S|}$, automatically respects a larger symmetry group $C(G)$ 
that additionally includes the bare logical algebra, which is a projective representation of $\mathbb{Z}_2^{\times |L|}$. Pairs of bare logical operators that define logical qubits anticommute, as an aside for readers familiar with the concept the resulting anticommutation relations can be taken to define a type of mixed anomaly valued in $H^2(\mathbb{Z}_2^{\times |L|},U(1))$ associated to the projective representation of the bare logical algebra~\cite{Tachikawa2020}. 

For topological subsystem codes in 2D~\cite{Bombin10}, the stabilizer symmetry is in fact a 1-form symmetry~\footnote{A $k$-form symmetry is generated by operators acting on submanifolds of codimension $k$.} associated to string operators of nontrivial pointlike superselection sectors~\cite{Gaiotto2015}. In this case the anomaly corresponds to the topological braiding phases of these pointlike sectors. In dimensions greater than two, the stabilizer symmetry of a topological subsystem code may be a higher $k$-form symmetry~\cite{Bombin2015,Kubica2015,Kubica2021}, or a more unconventional symmetry related to fracton topological order~\cite{Devakul2020b}. 


The time evolution $e^{-iHt}$ generated by a local Hamiltonian that commutes with a nontrivial stabilizer subsystem code symmetry, and hence takes the form introduced in Eq.~\eqref{eq:GaugeOpHam}, necessarily preserves a quantum memory as it commutes with the bare logical operators and therefore acts trivially within the $\mathcal{H}_L$ subsystems. 
More generally, any dynamical evolution generated by local terms that commute with the type of stabilizer subsystem symmetry we consider, and are hence linear combinations of gauge operators, preserve a quantum memory on $\mathcal{H}_L$. This includes open-system dynamics, which can be induced by unitary dynamics on an extended Hilbert space that commutes with the stabilizer symmetry on the original degrees of freedom.

For a more precise formulation of this point, consider a family of quantum channels describing time evolution parameterized by $t$, 
\begin{align}
\label{eq:GaugeChannel}
    \mathcal{E}_t(\rho) = \sum_{i} K_i(t)\, \rho \, K_i^\dagger(t) \, ,
\end{align}
where the Kraus operators $K_i(t)$ are (linear combinations of) gauge operators that satisfy the trace preserving condition $\sum_{i} K_i^\dagger(t) K_i(t) = \mathbbm{1}$. 
For the simple case of closed-system unitary evolution discussed above we have only a single Kraus operator $K(t)= e^{-iH t}$ for $H$ a local Hamiltonian of the form described in Eq.~\eqref{eq:GaugeOpHam}. 
This description also applies more generally, including in the case of monitored dynamics involving quantum measurements, in which case some Kraus operators may take the form of projections onto the eigenvalues of gauge operators, e.g., $K_i(t)=\frac{1}{2}(\mathbbm{1}-g_i)$ for a gauge operator $g_i$. 
In fact, the quantum memory we are discussing exists even at the level of individual quantum trajectories, where projective measurements are made without averaging over the measurement outcomes.

Any dynamics generated by local terms that respect a nontrivial stabilizer subsystem code symmetry (that corrects local errors) must take the form introduced in Eq.~\eqref{eq:GaugeChannel}. 
Hence the symmetry condition implies that the dynamics preserves a quantum memory, i.e., commutes with the bare logical algebra. 
The preservation of the quantum memory can be understood as a consequence of the mixed anomaly associated to the stabilizer subsystem code symmetry, see above. 

We emphasize that, in fixed stabilizer eigenspaces, the dynamics within the junk subsystem $\mathcal H_J$ can be totally generic or chaotic, or alternatively may be integrable such as when only a commuting set of gauge operators enters the dynamics, or in the simple case of conventional stabilizer codes where $\mathcal H_J$ is trivial.


\subsection{Adding perturbations}
\label{sec:Adding perturbations}

In the presence of Hamiltonian perturbations and couplings to the environment that do not respect the stabilizer symmetry, the dynamics generated by a stabilizer subsystem code may no longer protect $\mathcal H_{\rm L}$ in Eq.~\eqref{eq:SubsystemCodeBlockDecomposition} as a noiseless subsystem. 
To maintain protection of the quantum information in these subsystems, the most systematic approach is the implementation of quantum error correction, either through active measures or by passively relying on a sufficiently low temperature environment~\cite{Dennis2001,Bacon2005a}. This is an interesting direction with many facets that we leave to future work. Here we restrict our focus to passive extension of quantum memory lifetimes under closed-system unitary dynamics in the presence of nonsymmetric Hamiltonian perturbations, which may correspond to control errors. We rely on recently established results on prethermal quantum memories proved in Ref.~\cite{Else17a,Else20}. 

The results of Ref.~\cite{Else17a,Else20} apply to dynamics generated by Hamiltonians of the form
\begin{align}
\label{eq:PerturbedHam}
    H = H_0 + g V + \nu_1 \Sigma_1 + \dots + \nu_K \Sigma_K \, ,
\end{align}
where $H_0$ is a local Hamiltonian that obeys a U(1)$^{\times K}$ symmetry generated by the operators $\Sigma_k$ ($k=1,\dots,K$), and $V$ is a sum of symmetry-breaking local perturbations [with O(1) operator norm] with $g$ sufficiently small compared to all the $\nu_i$.
Note that both $H_0$ and $V$ generically contain non-commuting local terms, and that $H_0$ can be nonintegrable.
In order for the prethermalization arguments to hold, the U(1) generators $\Sigma_k$ must be sums of commuting local terms---in Sec.~\ref{sec:Bacon-Shor} we consider an example of a stabilizer subsystem code to which these prethermalization arguments cannot be applied due to nonlocality of the $\Sigma_k$. We demonstrate below that a sufficient condition guaranteeing that the $\Sigma_k$ are sums of commuting local terms is for the underlying subsystem code to be a topological subsystem code.
In the context of quantum memory, the crucial U(1)$^{\times K}$ symmetry is in fact closely related to a pair of stabilizer (subsystem) codes that we refer to as the \textit{local} and \textit{global} codes, which we now explain. 
We refer back to the example in Sec.~\ref{sec:Z2code} throughout the more general explanation. 

First, we specify the local subsystem code via a choice of stabilizer generators $S=\langle\{s_i\}\rangle$. As the name suggests, our focus is on cases where these generators have local support on a spatial lattice for practical reasons~\footnote{However, we consider at least one example where this is not the case.}. 
These stabilizer generators are given by star and plaquette terms for the example in Sec.~\ref{sec:Z2code}. 
There may be relations (i.e., redundancies) among these generators; we label a generating set of such relations as $\{r_j\}$, where the relations $r_j$ are defined such that
\begin{align}
\label{eq:StabilizerRelations}
\prod_{i\in r_j} s_i = \mathbbm{1} \, . 
\end{align}
In the example introduced in Sec.~\ref{sec:Z2code} there are no such relations.
Next, we define the group of gauge operators $G$ to be generated by all local operators contained in $C(S)$. 
Finally, the set of bare logical operators for the local code is given by $L=C(G)/S$. We demand that all elements of $L$ have macroscopic weight (if any were local those would already be included in $G$). We assume that the dimension $2^{|L|/2}$ of the logical algebra is at least two, otherwise there would be no quantum memory even for unperturbed dynamics. 
The logical operators are macroscopic string operators in the example covered in Sec.~\ref{sec:Z2code}. 
The local stabilizer subsystem code defines a block decomposition of the Hilbert space according to the eigenvalues of the stabilizer generators $\frac{1}{2}(\mathbbm{1}-s_i)={\lambda_i}$, see Eq.~\eqref{eq:SubsystemCodeBlockDecomposition}. 
Here, there may be nontrivial relations $r_j$ due to the the choice of generators being overcomplete. 
In this case, the blocks in Eq.~\eqref{eq:SubsystemCodeBlockDecomposition} labelled by eigenvalues not satisfying the relation conditions in Eq.~\eqref{eq:StabilizerRelations} are empty. 
In the example introduced in Sec.~\ref{sec:Z2code} these sectors correspond to configurations of pinned $e$ and $m$ anyons in the surface code. 

We now move on to define the U(1)$^{\times K}$ symmetry, and the global stabilizer subsystem code, in terms of the stabilizers of the local subsystem code. 
As above, each stabilizer generator defines an occupation number $\lambda_i\in\{0,1\}$ via the eigenvalue of $\frac{1}{2}(\mathbbm{1}-s_i)$. Any subset of stabilizer generators $S_k\subseteq \{ s_i \}$ (that involves an increasing number of generators as the number of physical qubits grows) defines a generator of a U(1) symmetry via 
\begin{align}
    \Sigma_k := \sum_{s_i \in S_k} \frac{1}{2}(\mathbbm{1}-s_i) \, .
\end{align}
A sufficient condition that guarantees the locality of the $\Sigma_k$ defined above is for the underlying local stabilizer subsystem code to be a topological subsystem code. In such codes, the stabilizers $s_i$ are local and mutually commuting by definition (see Sec.~\ref{sec:Stabilizer subsystem codes}).
Conversely, a U(1) symmetry generated by a sum of mutually commuting local Pauli terms always defines a topological stabilizer subgroup generated by those terms.
The local code in Sec.~\ref{sec:Z2code} is simply the topological surface code, and the U(1)$\times$U(1)  symmetry corresponds to $e$ and $m$ anyon number conservation. 

The U(1) symmetry generated by $\Sigma_k$ enforces conservation of the total occupation number,  $\sum \lambda_i^{(k)}$, of the stabilizers included in the set $S_k$. 
This induces a block decomposition of the Hilbert space,
\begin{align}
    \bigoplus_{\{n_k\}} \Big(\mathcal{H}_L \otimes \bigoplus_{\{ \lambda_i\} : \sum \lambda_i^{(k)} = n_k}{\mathcal{H}_J}_{\{ \lambda_i\}} \Big)_{\{n_k\}} \, ,
\end{align}
which is strictly coarser than the decomposition induced by the local stabilizer code symmetry, introduced in Eq.~\eqref{eq:SubsystemCodeBlockDecomposition}. In particular, all blocks of the local stabilizer symmetry that satisfy $\sum \lambda_i^{(k)}=n_k$, for the stabilizer generators in the set $S_k$, are collected into the single $\{ n_k\}$ block of the U(1)$^{\times K}$ symmetry. 
In the surface code example from Sec.~\ref{sec:Z2code} the blocks correspond to states with fixed $e$ and $m$ anyon number, collecting many different configurations with the same numbers of $e$ and $m$ anyons. 
The gauge operators of the local stabilizer code remain symmetric under the U(1)$^{\times K}$ symmetry. In addition, arbitrary local operators can be projected onto fixed occupation number eigenspaces of the overlapping local stabilizers $\frac{1}{2}(\mathbbm{1}-s_i)$ to produce further symmetric local operators. For example, the gauge generators of the global stabilizer code introduced below can be projected in this way to create symmetric local operators, as was done for the example in section~\ref{sec:Prethermalization}. 

Finally, we move on to define the global stabilizer subsystem code. 
Each set of stabilizers $S_k$ introduced above, that \textit{is not} a stabilizer relation, defines a nontrivial $\mathbb{Z}_2$ subgroup of both the U(1) symmetry and the local stabilizer symmetry. This subgroup is generated by a $\pi$ rotation of the form 
\begin{align}
e^{i \pi \Sigma_k}=\prod_{i\in S_k} s_i \, ,
\end{align}
which enforces conservation mod 2 of the occupation number of stabilizers in $S_k$. 
If $S_k$ is a relation, the above rotation by $\pi$ is trivial i.e. $e^{i \pi \Sigma_k}=\mathbbm{1}$ and the conservation of the occupation of stabilizers in $S_k$ mod  2 is a constraint (also known as a materialized symmetry~\cite{Kitaev2003,Brown2019}). 
For example, consider the toric code with periodic boundary conditions~\cite{Kitaev2003}; in this case, the global stabilizers $S_X$ and $S_Z$, consisting of the product of all star and plaquette stabilizers, respectively, are both relations (i.e., $S_X=S_Z=\mathbbm{1}$). The global stabilizer code is thus trivial (i.e., has trivial generators) and hence does not protect a quantum memory, even under dynamics that respect the U(1)$\times$U(1) symmetry corresponding to conservation of the total number of ``electric" and ``magnetic" anyons. To see this, note that an anyon can move around a noncontractible cycle of the torus, inducing a nontrivial logical action, via local steps that preserve the U(1)$\times$U(1) symmetry~\cite{Else17a}. This is no longer possible with the surface-code-like open boundary conditions considered in Section~\ref{sec:Z2code}. 
There the products over all star and plaquette operators, respectively, correspond to nontrivial boundary operators.

In the nontrivial case where $S_k$ is not a relation, the global stabilizer group is generated by the operators $\{e^{i \pi \Sigma_k}\}$ which induce a block decomposition 
\begin{align}
    \bigoplus_{\{ \sigma_j\} } \bigg(\mathcal{H}_L \otimes \bigoplus_{\{ \lambda_i\} : \sum \lambda_i^{(j)} = \sigma_j  } {\mathcal{H}_J}_{\{ \lambda_i\}} \bigg)_{\{ \sigma_j\}} \, ,
\end{align}
where $\sum \lambda_i^{(j)} = \sigma_j$ is only required to hold mod 2, with the sum taken over the set $S_j$ and with $\sigma_j=\frac{1}{2}(\mathbbm{1}-e^{i \pi \Sigma_j})$ the eigenvalue of the global stabilizer. 
In the example from Sec.~\ref{sec:Z2code} the block decomposition corresponds to the conservation of both $e$ and $m$ parity. 
Once again we define the the gauge group to be generated by all local operators that commute with all global stabilizers. This group contains all the gauge generators of the local stabilizer code. The bare logical operators are then given by the commutant of the gauge group, up to multiplication with stabilizers. Each bare logical must also be a bare logical of the local stabilizer code. The dressed logical operators for the global stabilizer code may be significantly lower weight than those of the local stabilizer code, due to the enlarged gauge group. In particular, the dressed logical operators may have constant weight, although by definition they must remain nonlocal otherwise they would have been included in the gauge group. 
The global stabilizer code may only be an error detecting code, even when the local stabilizer code is an error correcting code. However this is sufficient to preserve a quantum memory in a noiseless subsystem. 
For the example in Sec.~\ref{sec:Z2code} the gauge generators include single site Pauli operators in the bulk, and two body Pauli operators along the boundary. The dressed logical operators become two body on a pair of well separated boundaries and the global code is simply an error detecting code for certain single qubit errors on the boundary. 
For the examples of subsystem codes considered in Sec.~\ref{sec:Examples}, the global stabilizers act exclusively on the boundaries and indeed define single-Pauli error detecting codes there. 
In general, the global code should be a nontrivial error detecting code with no spatially local (dressed) logical operators. This suffices to ensure the persistence of a prethermal quantum memory under the dynamical evolution with U(1)$^{\times K}$ symmetry breaking perturbations introduced in Eq.~\eqref{eq:PerturbedHam}.


As discussed in section~\ref{sec:SubsystemCodeDynamics}, dynamics generated by a Hamiltonian that is a sum of gauge generators for the global subsystem code preserves a quantum memory associated to the noiseless subsystem $\mathcal{H}_L$. However, in the presence of generic local Hamiltonian perturbations we cannot provide a guarantee that this memory persists. 
For the stricter setting of dynamics generated by a local Hamiltonian obeying the full U(1)$^{\times K}$ symmetry we can apply the results of Refs.~\cite{Else17a,Else20} to ensure the existence of a prethermal memory time $t_*$ that is a stretched exponential in a parameter that measures the relative perturbation strength in a Hamiltonian of the form~\eqref{eq:PerturbedHam}. 
%
The results of Refs.~\cite{Else17a,Else20} imply that such a Hamiltonian $H$ can be expanded in a U(1)$^{\times K}$-symmetric perturbation series that is convergent up to an order proportional to $(\nu/g)^{1/K}$, where $\nu\equiv\sqrt{\sum^K_{i=1}\nu^2_i}$. This in turn implies that the evolution $e^{-itH}$ generated by the above Hamiltonian retains the U(1)$^{\times K}$ symmetry, up to small corrections, until a time
\begin{align}
    t_* \gtrsim e^{C(\nu/g)^{1/K}},
\end{align}
where $C>0$ is a constant.
In particular this implies that the perturbed evolution commutes with the global stabilizer symmetry up to small corrections, which in turn implies the preservation of a quantum memory associated to a noiseless subsystem. 
We may also choose to strictly enforce some subset of stabilizer symmetries, while enforcing others via a $U(1)^{\times K'}$ symmetry that is then allowed to be broken by sufficiently weak perturbations and similar results follow~\cite{Else17a,Else20}. 



\subsection{Further examples}
\label{sec:Examples}

The formalism outlined in Secs.~\ref{sec:Noiseless subsystems}--\ref{sec:Adding perturbations} yields a simple recipe for defining nontrivial quantum dynamics that preserves quantum information encoded in the initial state. In the most general case, any local quantum circuit composed of unitary gates generated by local products of the gauge generators $g_i\in G$ will preserve information encoded in the logical subsystem $\mathcal H_L$. Furthermore, one can allow hybrid unitary-projective circuits in which unitary dynamics is interspersed with projective local measurements, provided that the measurement operators are also local products of gauge generators. Alternatively, one can consider dynamics generated by a local Hamiltonian consisting of a sum over local products of gauge generators. Such models (and perturbations thereof) are studied in Secs.~\ref{sec:Z2Model} and \ref{sec:Prethermalization}. 

Regardless of their nature (i.e., whether they are Hamiltonian, unitary, or unitary-projective), the dynamics constructed in this way possess a symmetry group equal to the stabilizer group $S$. When the generators of the stabilizer group act on all lattice sites at once, the dynamics preserve a global symmetry, e.g. the $\mathbb Z_2\times\mathbb Z_2$ global symmetry of Sec.~\ref{sec:Z2code}, that protects the encoded quantum memory. Depending on the structure of the stabilizer group, the protecting symmetry can be a global, subsystem, or higher-form symmetry~\cite{Gaiotto2015,Vijay2016,Williamson2016}.

In this section, we consider several examples of such families of quantum dynamics and elucidate their essential features. We present in Appendix~\ref{App:Additional examples} a high-level discussion of other examples based on a variety of stabilizer codes in three or fewer spatial dimensions.  



\subsubsection{Bacon-Shor Code}
\label{sec:Bacon-Shor}
\begin{figure*}[t!]
\includegraphics[width=1.00\textwidth]{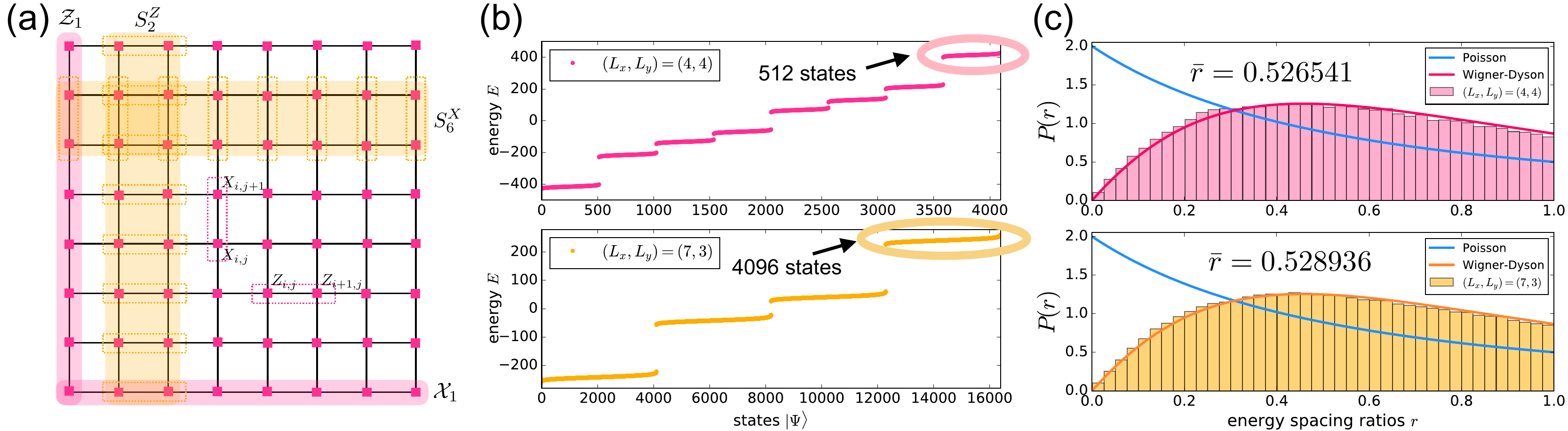}
\caption{
{\it The Bacon-Shor model.} 
        (a) Qubits (magenta squares) living on a square lattice with gauge generators $X_{i,j}X_{i,j+1}$ and $Z_{i,j}Z_{i+1,j}$ highlighted by dashed rectangles. Examples of the stabilizer generators $S^X_j$ and $S^Z_i$, i.e. $S^X_{6}$ and $S^Z_{2}$, are highlighted in yellow, and the logical operators $\mathcal Z_1$ and $\mathcal X_1$ each being a product of $Z(X)$-operators along a vertical (horizontal) path are highlighted in magenta. 
        (b) The energy spectra for two systems of size $(L_x,L_y) = (4,4)$ [top panel] and $(7,3)$ [bottom panel] containing a total of $16,384$ and $4,096$ states, respectively, are shown. The spectrum splits into eight (four) parts for $(L_x,L_y) = (4,4),(7,3)$ each separated by a relatively large energy difference. A single sector of the spectrum contains 512 states for $(L_x,L_y) = (4,4)$ and 4,096 states for a $(7,3)$-system. 
        (c) The histograms of energy level spacing ratios $r$ are  depicted again for the systems of size $(4,4)$ [top panel] and $(7,3)$ [bottom panel]. In order to lift any accidental degeneracies, mild disorder in the two bond couplings $J^x_{ij}$ and $J^z_{ij}$ is employed by uniformly drawing  $J^{x}_{ij}, J^{z}_{ij} \in [0.9,1.1] \, \forall\, i,j$ with $1 \leq i < L_{x}-1$ and $1 \leq j < L_{y}-1$. 
        To ensure convergence we averaged over 1,000 independent realizations of disorder for the system of size $(4,4)$ and over 200 disorder realizations for the $(7,3)$-system. 
        The resulting histogram data closely resemble a Gaussian  orthogonal  ensemble   (GOE)  distribution as expected. 
        The averaged r-values are $\bar r = 0.5265$ and  $\bar r = 0.5289$ for $(L_x,L_y) = (4,4)$ and $(7,3)$, respectively. 
}
\label{fig:baconshor}
\end{figure*}
\medskip
Consider a system of qubits on the vertices $\bm r=(i,j)\in\mathbb Z^2$ of an $L_x\times L_y$ square lattice as shown in Fig.~\ref{fig:baconshor}(a). We can define a Hamiltonian based on the Bacon-Shor code~\cite{Bacon2005a,Shor95} as follows:
\begin{align}
\label{eq:HBS}
    H_{\text{B-S}} = \sum^{L_x}_{i=1}\sum^{L_y}_{j=1}\left(J^x_{ij}\, X_{i,j}X_{i,j+1} +J^z_{ij}\, Z_{i,j}Z_{i+1,j}  \right).
\end{align}
The operators $X_{i,j}X_{i,j+1}$ and $Z_{i,j}Z_{i+1,j}$ generate the gauge group $G$ of the Bacon-Shor code.
The Hamiltonian \eqref{eq:HBS} has an extensive number of $\mathbb Z_2$ symmetries generated by the operators
\begin{subequations}
\label{eq:BaconShor-Stabilizer}
\begin{align}
    S^X_j=\prod^{L_x}_{i=1}X_{i,j}X_{i,j+1},\indent \indent 1\leq j <L_y
\end{align}
and
\begin{align}
    S^Z_i=\prod^{L_y}_{j=1}Z_{i,j}Z_{i+1,j},\indent \indent 1\leq i <L_x,
\end{align}
\end{subequations}
for a total of $L_x+L_y-2$ $\mathbb Z_2$ symmetries (see Fig.~\ref{fig:baconshor}(a)).
The operators \eqref{eq:BaconShor-Stabilizer} generate the stabilizer group $S$ of the Bacon-Shor code.
Physically, $S^X_j$ and $S^Z_i$ generate a \textit{subsystem symmetry}, i.e., one that acts on rigid one-dimensional submanifolds of the two-dimensional space.
This model has one encoded qubit with bare logical operators
\begin{align}
    \mathcal X_j &= \prod^{L_x}_{i=1}X_{i,j},\indent \indent \mathcal Z_i = \prod^{L_y}_{j=1}Z_{i,j}.
\end{align}
Note that, for any $j,j'$, $\mathcal X_j$ can be transformed into $\mathcal X_{j'}$ by multiplication by some number of $S^X_j$, and similarly for any two $\mathcal Z_i$ and $\mathcal Z_{i'}$. Thus, there is only one encoded logical qubit in this model.

We now study the Bacon-Shor Hamiltonian~\eqref{eq:HBS} 
to illustrate that, like the model studied in Sec.~\ref{sec:Z2Model}, it is quantum chaotic when all symmetries and the state of the logical qubit are resolved.
To do this, we again study the statistics of energy levels within a symmetry sector. We aim to resolve the extensive number of $\mathbb Z_2$ symmetries generated by 
\eqref{eq:BaconShor-Stabilizer}. The resolution of the $(L_x-1)$ operators $S^Z_i$ for $1\leq i <L_x$ is trivial in the convenient $Z$-basis. 
Further we circumvent the twofold energy eigenvalue degeneracy brought on by the single encoded qubit by taking into account the two possible quantum numbers of a single $\mathcal Z_i$ for an arbitrarily chosen $i$, i.e. $\mathcal Z_{1}$ as depicted in Fig.~\ref{fig:baconshor}(a). 
The resolution of these symmetries reduces the size of the Hilbert space significantly. The full Hilbert space of $N=L_x L_y$ qubits, dim$(\tilde{H}_{B-S}) = 2^{L_x L_y}$, shrinks 
to dim$(\tilde{H}_{B-S}|\{S^{Z}_{i}\},\mathcal Z_{i=1})$ $= 2^{L_x L_y-L_x}$. 

To resolve the symmetries generated by $S^{X}_i$, $1\leq i < L_y$, which are not diagonal in the $Z$-basis, we proceed by modifying the Hamiltonian 
\eqref{eq:HBS} by adding these operators with large prefactors $\mu_i$. 
The new Hamiltonian reads 
\begin{align}
\label{eq:HBS1}
    H = H_{\text{B-S}} 
    + \sum^{L_y-1}_{i=1} \mu_{i} S^X_i 
\end{align}
with $\mu_{i} = 10\left(1, \sqrt{2}, \sqrt{3}, \sqrt{5}, \sqrt{6}, \sqrt{ 7}, \ldots\right)$ for $i = 1, \ldots,L_y-1$. The $\mu_i$ values are chosen to be irrationally related to one another in order to guarantee that energy levels from each sector labeled by eigenvalues of the $S^{X}_i$ are separated by sizeable energy gaps. To avoid accidental degeneracies we introduce a disordered set of bond couplings  
$J^{x}_{ij}, J^{z}_{ij} \in [0.9,1.1] \, \forall\, i,j$ with $1 \leq i < L_{x}-1$ and $1 \leq j < L_{y}-1$. 
We present results for two different systems of sizes 
$(L_x,L_y)=(7,3)$ and $(4,4)$, the former resulting in eight energy-separated sectors with each sector containing $512$ states while the latter consists of four energy sectors with each sector harboring $4,096$ states [see Fig.~\ref{fig:baconshor}.(b)].

To emphasize the nonintegrable nature of the model, we analyzed the level statistics for the two different system sizes mentioned above. 
To ensure convergence, the histograms shown in Fig.~\ref{fig:baconshor}(c) contain data from a large set of independent numerical simulations with different realizations of the random couplings defined above. We used $1,000$ independent realizations for the $(4,4)$-system and $200$ independent realizations for the $(7,3)$-system. 
We observe that the histograms are in very good agreement with the Wigner-Dyson distributions. The average values for the level spacing ratio $r$ are found to be ${\bar r = 0.5265}$ and $\bar r = 0.5289$ for systems sizes ${(L_x,L_y)=(4,4)}$ and $(7,3)$, respectively. 
This is quite close to the exact value of 
$\bar r_{\rm GOE} \approx 0.536$ belonging to the Gaussian orthogonal ensemble (GOE), as expected for generic real matrices and markedly different from the corresponding value $\bar r_{\rm Poisson} \approx 0.386$ for the Poisson distribution~\cite{Atas13}. 

The global error detecting code contained within Bacon-Shor is essentially the same as that in our main example of the surface code discussed in Sec.~\ref{sec:Z2code}, which we discuss further in Sec.~\ref{sec:SurfCodeExample} below. The relevant global subgroup of the stabilizer group $S$ is generated by the products 
\begin{align}
    S_X=\prod_{j=1}^{L_y-1} S_j^X \, ,
    &&
    S_Z = \prod_{i=1}^{L_x-1} S_i^Z \, .
\end{align}
As in the surface code example, these operators have support only along the perimeter of the lattice, and the gauge group of the global code is generated by all local Pauli operators that commute with $S_X$ and $S_Z$.
However, in this case the global stabilizer generators cannot be written as products of local, mutually commuting terms, unlike the case of the surface code discussed in Sec.~\ref{sec:Z2code}. This can be viewed as a consequence of an anomaly of the symmetry generated by $S^X_j$ and $S^Z_i$, albeit one slightly different from the anomaly discussed in Sec.~\ref{sec:SubsystemCodeDynamics} that guarantees the existence of a nontrivial logical algebra. This anomaly guarantees that one cannot write a Hamiltonian of the form in Eq.~\eqref{eq:PerturbedHam} with locally generated $\Sigma_k$ unless additional degrees of freedom are added.


To see how the anomaly precludes the application of the prethermalization arguments, it is informative to attempt to write down a set of local, commuting U(1) generators $\Sigma_k$ like those appearing in Eq.~\eqref{eq:PerturbedHam}. To do this, it is natural to rewrite
\begin{align}
\begin{split}
    S^X_j &\propto \exp\bigg(-i\frac{\pi}{2}\sum^{L_x}_{i=1}X_{i,j}X_{i,j+1}\bigg) \\
    S^Z_i &\propto \exp\bigg(-i\frac{\pi}{2}\sum^{L_y}_{j=1}Z_{i,j}Z_{i+1,j}\bigg),
\end{split}
\end{align}
where the proportionality omits a global phase. It is then natural to identify local U(1) generators
\begin{align}
\begin{split}
    \Sigma^X_j &= \frac{1}{2}\sum^{L_x}_{i=1}(\mathbbm 1 - X_{i,j}X_{i,j+1})\\
    \Sigma^Z_i &= \frac{1}{2}\sum^{L_y}_{j=1}(\mathbbm 1 - Z_{i,j}Z_{i+1,j}).
\end{split}
\end{align}
However, these U(1) generators manifestly fail to commute with one another, as expected from the above discussion of the anomaly. In fact, they fail to commute with the Hamiltonian \eqref{eq:HBS}, so there is no basis to apply the prethermalization arguments of Refs.~\cite{Else17a,Else20}. One can, however, define a pair of genuine, albeit nonlocal, U(1) symmetry generators, namely
\begin{align}
\begin{split}
    \Sigma^X &= \frac{1}{2}\sum^{L_y-1}_{j=1}(\mathbbm
    1 - S^X_j)\\
    \Sigma^Z &= \frac{1}{2}\sum^{L_x-1}_{i=1}(\mathbbm 1 - S^Z_i).
\end{split}
\end{align}
These generators are mutually commuting, but they are nonlocal. This also precludes the application of prethermalization arguments, since they require local Hamiltonians as input~\cite{Abanin17,Else17a,Else20}.

\subsubsection{Surface code and $\mathbb{Z}_2$ gauge theory}
\label{sec:SurfCodeExample}


The surface code is defined on a square lattice with rough top and bottom boundaries and smooth left and right boundaries, with one qubit placed on each link $\ell$ [see Fig.~\ref{fig:histo_entanglement}(a)]. Its stabilizer generators are 
\begin{align}
    A_v = \prod_{\ell\ni v} X_\ell \, , && B_p = \prod_{\ell\in p} Z_\ell \, 
\end{align}
which correspond to weight-four operators in the bulk and weight-three operators on the boundary. 
The product of all vertex and plaquette terms generate a global $\mathbb{Z}_2 \times \mathbb{Z}_2$ stabilizer subgroup symmetry which is suitable for the U(1)$\times$U(1) construction outlined in Sec.~\ref{sec:Adding perturbations}. It is also possible to consider, e.g., the vertex terms as an unbroken gauge symmetry and use only the plaquette terms to generate a single U(1) symmetry.

\begin{figure}[t!]
\includegraphics[width=1.00\columnwidth]{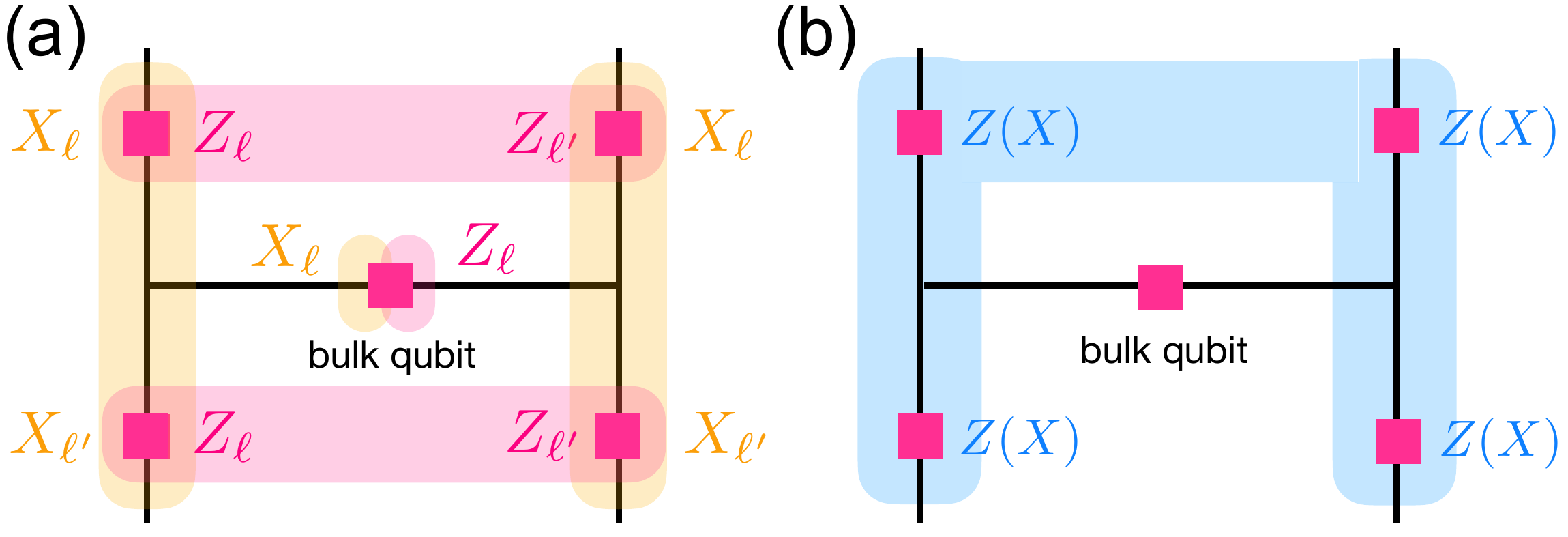}
\caption{
         {\it ``H''-shaped surface code lattice with five qubits.} 
         (a) The gauge generators consist of a single $Z_{\ell}$ ($X_{\ell}$)-operator on the bulk qubit together 
         with pairs of $Z_{\ell} Z_{\ell^{\prime}}$-operators on consecutive edges along the top and bottom boundaries (magenta shaded) and pairs of $X_{\ell} X_{\ell^{\prime}}$-operators on  
         consecutive edges along the left and right boundaries (orange shaded). 
         (b) The global code with stabilizers $Z Z Z Z$ ($X X X X$) (blue shaded). 
}
\label{fig:H_code}
\end{figure}

The global $\mathbb{Z}_2\times \mathbb{Z}_2$ stabilizer symmetry serves as an error detecting subsystem code for an odd number of $Z$ errors along the top or bottom boundaries, and similarly for an odd number of $X$ errors along the left or right boundaries. This code corresponds to the generators considered in section~\ref{sec: Disentangling unitary} in the disentangled picture. 
This is very similar to the subsystem boundary error detecting code derived from Bacon-Shor above. However, it contains extra qubits in the bulk that allow the global symmetries to be generated by products of local commuting terms, which is essential for the application of the U(1) construction. 
The symmetric local Hamiltonian terms can be viewed as gauge generators of this subsystem error detecting code. The simplest generating set corresponds to single $X_\ell,Z_\ell$ operators on qubits in the bulk together with pairs $Z_\ell Z_{\ell'}$ on consecutive edges along the top or bottom boundary and similarly $X_\ell X_{\ell'}$ on consecutive edges along the left or right boundary. A subset of these terms appears in the model discussed in Sec.~\ref{sec:Z2Model}.
The bare logical operators of the global code can be chosen as $\mathcal X = \prod X_\ell$ for links $\ell$ in the top rough boundary, and $\mathcal Z = \prod Z_\ell$ for links $\ell$ in the left smooth boundary.
There are also dressed logical operators of weight two given by a pair of $X$ ($Z$) operators on the left and right (top and bottom) boundaries. We remark that these operators are spatially nonlocal, and remain so under multiplication with gauge generators. 
This global stabilizer subsystem code is, in some sense, a generalization of the simplest error detecting code on four qubits, which has stabilizer generators $XXXX,ZZZZ$. In particular, on a small surface code lattice with five qubits on edges forming an ``H" (see Fig.~\ref{fig:H_code}),  the global code reduces to the four-qubit error detecting code for the boundary qubits, together with a single trivial bulk qubit whose $X$ and $Z$ operators are gauge generators. 

\subsubsection{Color code}
\begin{figure}[t!]
\includegraphics[width=1.00\columnwidth]{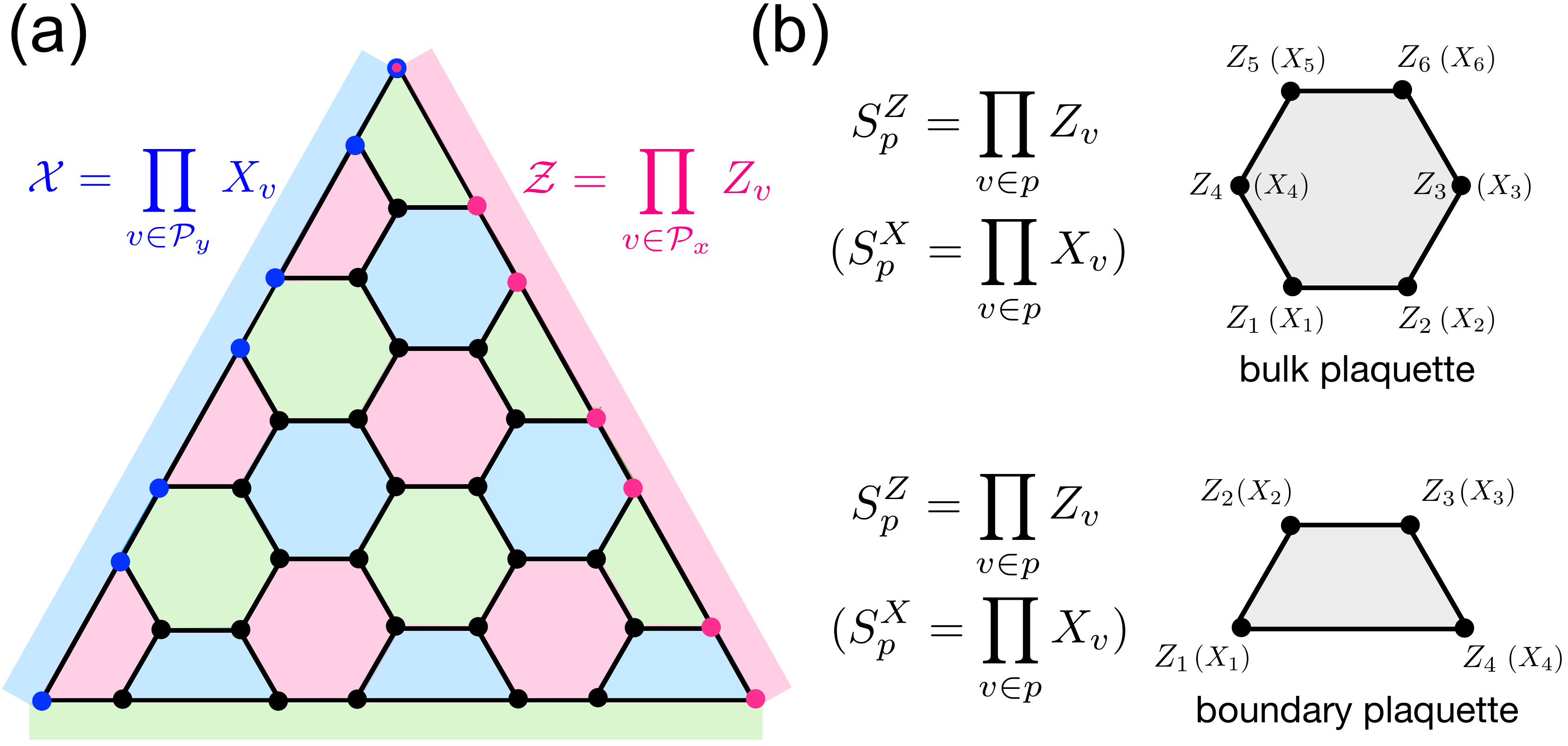}
\caption{
{\it The color code on the honeycomb lattice.} 
         (a) A triangular patch of the color code on the honeycomb 
         lattice. Plaquettes appear in three colors, i.e. blue, green, and red. The logical operators $\mathcal Z$ and $\mathcal X$ are given by product of $Z_\ell$'s and $X_\ell$'s along a 
         boundary path ${\mathcal P_x}$ and ${\mathcal P_y}$, respectively. 
         (b) The generators of the stabilizer group 
         $S_p^{Z} = \prod_{v \in p} Z_v$ and 
         $S_p^X = \prod_{v \in p} X_v$ 
         consisting each of six (four) 
         $Z(X)$-operators acting on the qubits of each hexagon 
         in the bulk (on the boundary) independent from its color.  
}
\label{fig:colorcode}
\end{figure}
The stabilizer group for the color code is generated by
\begin{align}
    S_p^X = \prod_{v \in p} X_v \, , &&  S_p^{Z} = \prod_{v \in p} Z_v \, ,
\end{align}
with qubits on vertices of a trivalent lattice with three-colorable plaquettes taken to be red, green, and blue. 
The code has four independent global relations generated by the product of $S_p^X$ on all red and green plaquettes and all green and blue plaquettes, and similarly for $S_p^Z$. 
To introduce a logical qubit, we consider a triangular patch of the honeycomb lattice with one boundary of each color, see Fig.~\ref{fig:colorcode}(a). 
The logical $X$ operator is given by the product of $X$ operators on physical qubits along a single boundary, i.e. $\mathcal X = \prod_{v\in\mathcal P_y} X_v$. 
Similarly, for the $Z$ logical operator we have
$\mathcal Z = \prod_{v\in\mathcal P_x} Z_v$ 
[see Fig.~\ref{fig:colorcode}(a)]. 
Under these boundary conditions, the global relations become a $\mathbb{Z}_2^{\times 4}$ global symmetry group along the boundary. 
This code is suitable for application of the prethermalization construction of Sec.~\ref{sec:Adding perturbations} by utilizing this global $\mathbb{Z}_2^{\times 4}$ symmetry to define a global U(1)$^{\times 4}$ symmetry. 
The relevant error detecting boundary subsystem code has gauge generators $X_v,Z_v,$ on all bulk qubits $v$, two body $X_vX_{v'},Z_vZ_{v'}$ gauge generators along all boundaries away from the corners, and three body $X$ and $Z$ terms at each corner, see Fig.~\ref{fig:colorcode}(a). The stabilizers are given by the global $\mathbb{Z}_2^{\times 4}$ boundary symmetry. 
There are weight-two dressed logical operators generated by a pair of Pauli $X$ ($Z$) operators on a corner and the opposite boundary. These dressed logicals are spatially nonlocal, and cannot be made local under multiplication with gauge generators.  
The smallest instance of the global subsystem error detecting code is defined on seven qubits, with a single trivial bulk qubit that has its $X$ and $Z$ operators included as gauge generators, together with six boundary qubits that have gauge generators 
\begin{align}
\begin{split}
    &XXIIIX, \quad IXXXII, \quad IIIXXX,
    \\
    &ZZIIIZ, \quad IZZZII, \quad IIIZZZ
\end{split}
\end{align} 
and 
stabilizer generators 
\begin{align}
    XIXXIX,\ XXIXXI,\ ZIZZIZ,\ ZZIZZI. 
\end{align}

\section{Conclusion}
\label{sec:conclusion}

In this work, we have shown that the formalism of noiseless subsystems and stabilizer subsystem codes can be used to define interesting classes of dynamics that preserve quantum information encoded in the initial state of a quantum many-body system. This formalism is sufficiently general to encompass a variety of quantum dynamics, including unitary and open-system dynamics. In the context of Hamiltonian systems, which are the focus of the examples considered in Secs.~\ref{sec:Z2Model} and \ref{sec:Bacon-Shor}, stabilizer subsystem codes can be used to define families of strongly nonintegrable Hamiltonians with encoded logical qubits. The dynamical quantum memory in the systems we consider is symmetry-protected in the sense that the lifetime of the encoded quantum information is infinite when the stabilizer symmetry of the subsystem code is strictly enforced. In the presence of perturbations that break the stabilizer symmetry, we argued that existing results on prethermalization can be leveraged to provide a parametrically long lifetime of the encoded quantum state, provided that an appropriate global error-detecting code within the parent subsystem code can be identified.

One intriguing direction for future work concerns the applicability of the ETH for Hamiltonians like the ones considered in Secs.~\ref{sec:Z2Model} and \ref{sec:Bacon-Shor}. The numerical results we have obtained for these Hamiltonians strongly suggest that these models obey the ETH once the stabilizer symmetry group and the state of the logical qubit are resolved. However, the ETH is a statement about matrix elements of local operators, so a study of the statistics of such matrix elements in eigenstates of these models must be performed in order to determine whether it holds in these models. In the surface code example presented in Sec.~\ref{sec:Z2Model}, it is reasonable to expect that, e.g., single-site Pauli operators in the bulk of the system will obey the ETH, as these operators commute with the stabilizer symmetry generators $\tilde{S}_X$ and $\tilde{S}_Z$. However, applying such operators on the boundary of the system can change the stabilizer symmetry sector and flip one or both of the logical operators. Since the model in Sec.~\ref{sec:Z2Model} has pervasive spectral degeneracies when the stabilizer symmetry is not resolved, it is possible that these boundary operators do not obey the ETH. The possibility that ETH may hold in the bulk but not on the boundaries of some systems is an enticing one that warrants further study.

Another avenue for further exploration is to consider the stability of other classes of dynamics based on subsystem codes. For example, one can consider Floquet or even random unitary circuits generated by gauge operators of a subsystem code and inquire whether the lifetime of the encoded quantum information under these circuits can be parametrically enhanced in the presence of symmetry-violating perturbations. One promising direction in these cases is to consider dynamical-decoupling-like protocols~\cite{Viola99,Tran21} using the stabilizer generators. 

It would also be interesting to consider the mechanisms discussed in this work in the setting of open or monitored quantum systems. For example, it would be worthwhile to study how perturbations affect the dynamics in the setting of hybrid unitary-projective circuits---for example, measuring stabilizer generators at a sufficient rate can suppress the propagation of errors, similar to what was observed for topological codes in Refs.~\cite{Sang21,Lavasani21a,Lavasani21b}. An additional question worth investigating is whether prethermalization could be invoked to protect the quantum memory when a system with a Hamiltonian of the form \eqref{eq:PerturbedHam} is coupled to a bath via a small term that does not respect the U(1)$^{\times K}$ symmetry.

Another related direction to explore is subsystem codes with dynamically generated logical qubits, which were recently defined in the context of Kitaev's honeycomb model~\cite{Hastings2021,Gidney2021}. In particular, one can consider a wider family of topological subsystem codes that may generate analogous dynamical memories. 
In all of the dynamical settings mentioned above, it would be interesting to understand the influence that different choices of qualitatively distinct subsystem codes have on the dynamics. This includes comparing the behaviors of topological and nontopological subsystem codes, or even different topological codes with qualitatively distinct properties, such as self-correction~\cite{Dennis2001,Bacon2005a,bravyi2013quantum}, single-shot error-correction~\cite{Bombin2015b,Brown2016}, or fractal symmetries~\cite{Haah2011,PhysRevLett.116.027202,Vijay2016,Williamson2016,PhysRevB.95.155133,Devakul2020b}.

Finally, we comment on the experimental realizability on present-day quantum hardware of the physics studied in this work. Recently, two groups have reported the implementation of repeated quantum error correction in a distance-three surface code~\cite{Krinner21,Zhao21}. To achieve this, both works perform measurements of the surface code stabilizers $\tilde{A}_v$ and $\tilde{B}_p$ using one ancilla qubit per vertex $v$ and plaquette $p$. This experimental setup is sufficient to implement quantum simulation of the surface-code-inspired Hamiltonian \eqref{eq:ModelHZ2}. For example, to realize the unitary gate $e^{i\theta \tilde{B}_p}$, one can use a sequence of CNOT gates with the ancilla qubit $p$ as the target, followed by a rotation $e^{i\theta Z_p}$ of the ancilla and the inverse of the preceding CNOT sequence (see, e.g., Sec.~4.7.3 of Ref.~\cite{Nielsen2010}). In fact, the necessary sequence of CNOTs is the same one needed to measure the stabilizer $\tilde{B}_p$. A similar circuit (up to single qubit rotations) can be used to realize the gate $e^{i\theta\tilde{A}_v}$. Combined with readily available single-qubit rotations around the $X$ and $Z$ axes, these ingredients are all that is needed to perform quantum simulation of Eq.~\eqref{eq:ModelHZ2}. One could then prepare a logical state of the surface code using either the methods of Refs.~\cite{Krinner21,Zhao21} or those of Ref.~\cite{Satzinger21}, and then evolve this state under the Hamiltonian~\eqref{eq:ModelHZ2}, or an appropriate Floquet version thereof, and measure the expectation values $\braket{\tilde{\mathcal X}}$ and $\braket{\tilde{\mathcal Z}}$ as functions of time. The evolution could be performed with and without the perturbations \eqref{eq:U1Pert} to deduce their effect on the logical qubit lifetime. Such a study would provide a nontrivial application of this surface code architecture even before the large-scale implementation of error correction on such devices.



\begin{acknowledgments}
T.I.~acknowledges the hospitality of the Aspen Center for Physics, which is supported by National Science Foundation grant PHY-1607611. 
D.W. acknowledges support from the Simons Foundation and related discussions with Arpit Dua, Sarang Gopalakrishnan, Michael Gullans, David Huse, Matteo Ippoliti, and Vedika Khemani.
\end{acknowledgments}


\appendix


\section{Encoding multiple qubits}
\label{sec: Encoding multiple qubits}
\begin{figure}[t!]
\includegraphics[width=1.00\columnwidth]{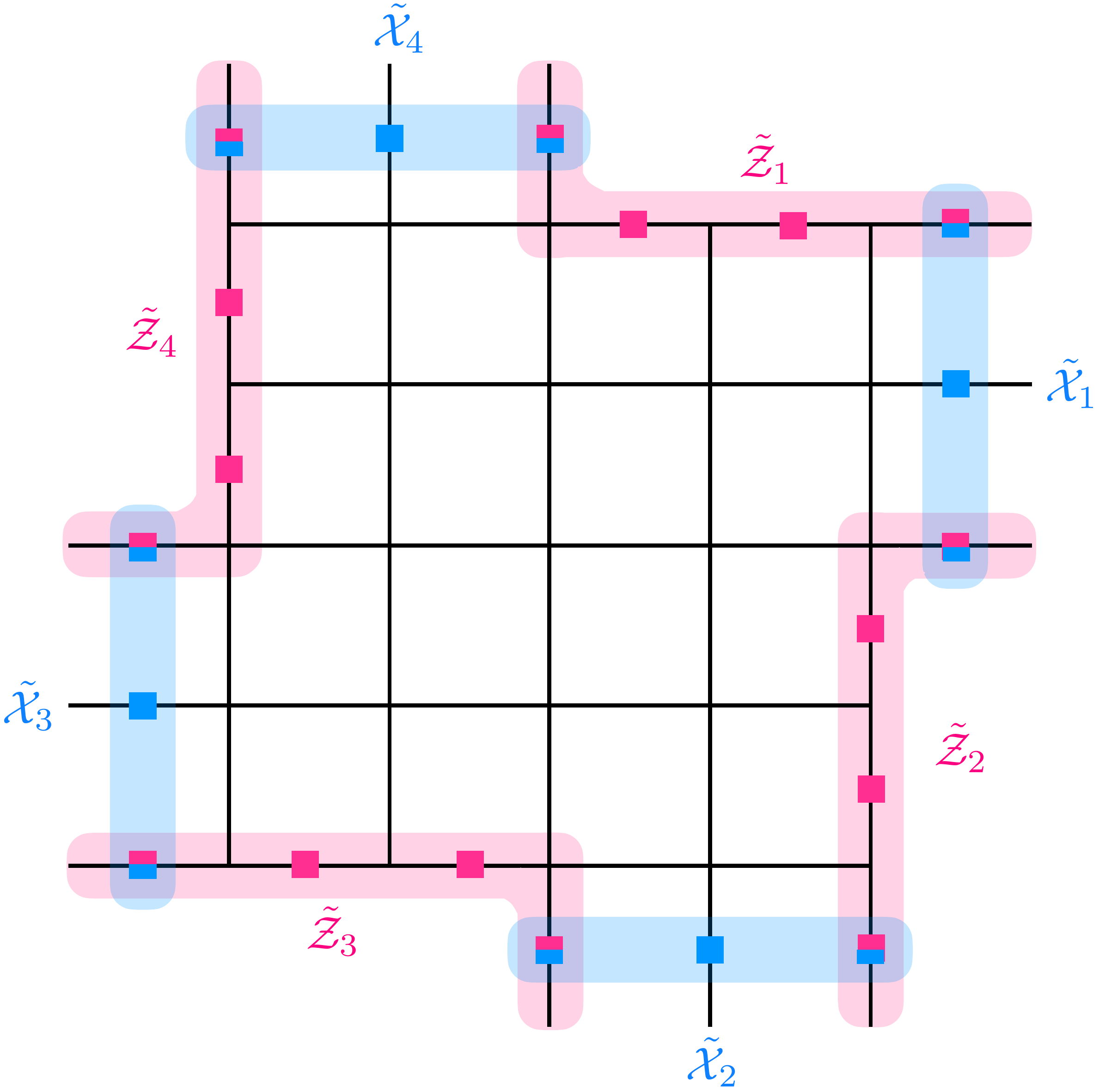}
\caption{
{\it Surface-code lattice encoding multiple qubits.} An example of a square lattice with alternating rough and smooth boundaries is shown. As in the main text, qubits live on the links of the lattice. There are four rough boundary segments and four smooth boundary segments, corresponding to $N=4$. Operators $\tilde{\mathcal Z}_i$ and $\tilde{\mathcal X}_i$ analogous to Eqs.~\eqref{eq:BareLogicals} are indicated by magenta and blue shaded regions, respectively. Each $\tilde{\mathcal Z}_i$ or $\tilde{\mathcal X}_i$ operator consists of the product of $Z$ or $X$ operators within the shaded region, respectively.
}
\label{fig:multiqubit}
\end{figure}

There exists an analog of the setup detailed in Secs.~\ref{sec:Setup}--\ref{sec: Encoded logical qubit and connection to subsystem codes} that encodes more than one logical qubit. To do this, one can define analogs of Eqs.~\eqref{eq:HilbertSpace}, \eqref{eq:GaugeGenerators}, and \eqref{eq:TopoSymm} for a square lattice with alternating smooth and rough boundaries (see Fig.~\ref{fig:multiqubit}). Then, one can pass to the disentangled picture in the manner described in Sec.~\ref{sec: Disentangling unitary}. Next, one can define operators analogous to the logical operators~\eqref{eq:BareLogicals}. For each smooth boundary, one can define operators $\tilde{\mathcal Z}_i$ for $i=1,\dots,N_{\rm smooth}$, where $N_{\rm smooth}$ is the number of smooth boundaries. Likewise, for each rough boundary, one can define operators $\tilde{\mathcal X}_j$ for $i=1,\dots,N_{\rm rough}$, where $N_{\rm rough}$ is the number of rough boundaries. Assuming for simplicity that $N_{\rm smooth}=N_{\rm rough}=N$, the topological symmetry generators \eqref{eq:TTopoSym} can then be written as $\tilde S_X=\prod^N_{j=1}\tilde{\mathcal X}_j$ and $\tilde S_Z=\prod^N_{i=1}\tilde{\mathcal Z}_i$. Upon restricting to specific eigenspaces of the conserved quantities $\tilde S_X$ and $\tilde S_Z$, this setup encodes $N-1$ logical qubits. An example of a commuting basis of conjugate logical operators for these qubits is
\begin{align}
    \begin{split}
        \tilde{\mathcal Z}_1,&\indent  \tilde{\mathcal X}_1\\
        \tilde{\mathcal Z}_1\tilde{\mathcal Z}_2,&\indent  \tilde{\mathcal X}_2\\
        &\vdots\\
        \prod^{N-1}_{i=1}\tilde{\mathcal Z}_i,&\indent \tilde{\mathcal X}_{N-1}.
    \end{split}
\end{align}
One can verify by explicit calculation that the operators on each line above anticommute with one another, while the operators in different lines commute with one another.
Note that, in order for this modified surface code to have macroscopic distance, we must fix $N$ to be constant while taking the thermodynamic limit on the lattice. This is because we need the length of each boundary segment to be macroscopic in order to avoid finite-size effects like those discussed around Eq.~\eqref{eq:finite-size lifetime}.

It is interesting to note that all logical qubits in this setup are protected by the same $\mathbb Z_2\times\mathbb Z_2$ symmetry. We therefore expect that the prethermalization analysis of Refs.~\cite{Else17a,Else20} would still hold upon promoting the $Z_2\times\mathbb Z_2$ symmetry to a U(1)$\times$U(1) symmetry and adding the appropriate fields. 

\section{Additional examples for dimensions $D \leq 3$}
\label{App:Additional examples}

Harnessing the framework of stabilizer subsystem codes with local generators leads to many further examples, some of which we briefly discuss below, organized by spatial dimension, i.e. one-dimensional (1D), two-dimensional (2D), and three-dimensional (3D) subsystem codes. 

    \subsection*{1D}
    The Kitaev wire~\cite{Kitaev01} is a Majorana fermion subsystem code with a pair of Majorana modes per site $\gamma^A_j,\gamma^B_j$. The local stabilizer generators are $i \gamma_j^B\gamma_{j+1}^A$, which couple neighboring sites. The logical operators are $\prod_j i \gamma_j^A \gamma_j^B$ and $\gamma_1^A$. This local code is capable of correcting local errors that obey global fermion parity symmetry. 
    The global subsystem code with stabilizer $\prod_j i \gamma_j^B\gamma_{j+1}^A$ is capable of detecting a local error affecting the boundary that obeys global fermion parity symmetry, which is utilized in the U(1) prethermalization construction of Ref.~\cite{Else17a}. 
    
\begin{figure}[t!]
\includegraphics[width=1.00\columnwidth]{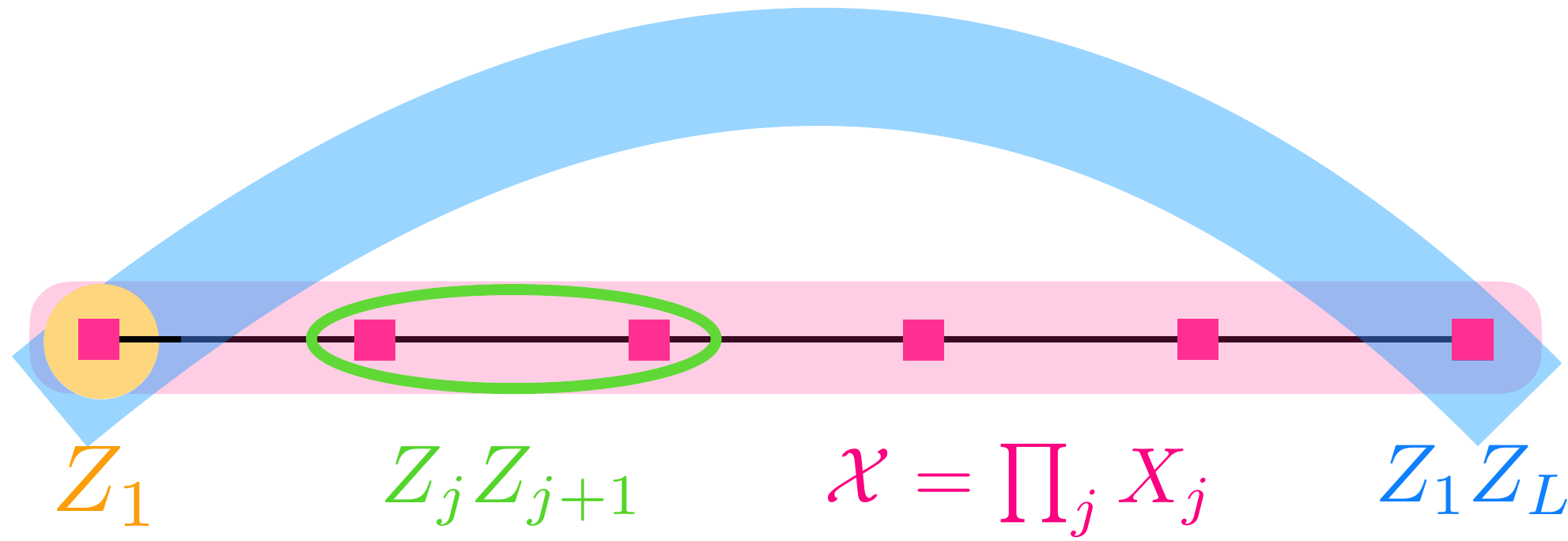}
\caption{
{\it The repetition code.} 
         The one-dimensional repetition code on a chain of length $L$. The green oval indicates the stabilizer generator $Z_{j}Z_{j+1}$ while the logical operators are $Z_{1}$ (orange circle) and $\mathcal X = \prod_{j=1}^{L} X_{j}$ (magenta shaded). 
         The global code has stabilizer 
         $\prod_{j=1}^{L} Z_{j}Z_{j+1} = Z_{1}Z_{L}$ (blue shaded). 
}
\label{fig:repetition}
\end{figure}

    The above example is related by a Jordan-Wigner transformation to the 1D repetition code, or Ising model with stabilizer generators $Z_jZ_{j+1}$. The logical operators are $Z_1$ and $\prod X_j$ (see Fig.\ref{fig:repetition}). This local code is capable of correcting local bit-flip (i.e., $X$) errors. The global code with stabilizer $\prod Z_jZ_{j+1} = Z_1 Z_L$ is capable of detecting a local bit flip error affecting the boundary spins. 
    Dynamics respecting the global code symmetry is generated by gauge operators $Z_jZ_{j+1}$ and $X_i$ on all sites excluding the boundary spins. 
    This global code was similarly harnessed for the U(1) prethermalization arguments of Ref.~\cite{Else17a}.

    In both of the 1D examples above, one of the logical operators corresponds to a global symmetry, implying that the quantum nature of the memory can only be probed by nonsymmetric initial states and operators. This poses a particular challenge in fermionic systems, where global fermion parity symmetry cannot be broken. This obstruction could be circumvented by coupling the system to an auxiliary fermionic system in such a way that fermion-parity-violating processes within the chain could be implemented while preserving the global fermion parity.
    
    \subsection*{2D}
    In addition to the topological stabilizer code examples in above subsections, there are topological stabilizer subsystem codes with nontrivial gauge degrees of freedom that can evolve under the dynamics generated by local gauge generators that commute with the topological 1-form stabilizer symmetry. 
    Examples include the subsystem surface code~\cite{Bravyi2013}, which obeys a $\mathbb{Z}_2\times\mathbb{Z}_2$ 1-form stabilizer symmetry with the braiding of the toric code anyons; the gauge color code~\cite{Bombin2009,Kargarian2010,Bombin10}, which obeys a $\mathbb{Z}_2\times\mathbb{Z}_2$ 1-form stabilizer symmetry with the braiding of the three-Fermion anyon theory; and further related models~\cite{Suchara11}. These examples demonstrate that directly enforcing a 1-form symmetry with a nondegenerate braiding can protect a quantum memory directly in 2D, despite hosting nontrivial dynamics. 
    
    \subsection*{3D}
    There are natural generalizations of the surface code and color code to three dimensions~\cite{Dennis2001,Bombin2007}. The stabilizer group of the surface code consists of a $\mathbb{Z}_2$ 1-form symmetry and a $\mathbb{Z}_2$ 2-form symmetry that mutually capture the braiding of the topological $\mathbb{Z}_2$ loop and pointlike excitations, respectively. Since the 1-form symmetry persists to finite temperatures~\cite{Castelnovo2007,Castelnovo2008}, Ref.~\cite{Else17a} noted that only the global symmetry coming from the product of all star terms (which measures the global parity of pointlike excitations) need be enforced to preserve a quantum memory at sufficiently low temperatures. 
    The color code is local unitary equivalent to copies of surface code~\cite{kubica2015unfolding} and so exhibits similar code properties. 
    In addition to the global symmetry stabilizer subgroups acting nontrivially on boundaries with topological boundary conditions, in these 3D codes it is also possible to introduce linelike topological defects along which the global symmetry may be supported~\cite{Else17a,else2017cheshire}. 

    There are subsystem versions of the surface code~\cite{Kubica2021} and color code~\cite{Bombin2015,Kubica2015} in 3D which support nontrivial gauge degrees of freedom. The stabilizers in these codes are 1-form symmetries, and give rise to global constraints that lead to boundary symmetries on open boundary conditions suitable for the application of the prethermalization construction of Sec.~\ref{sec:Adding perturbations}. Interestingly, these codes also display single-shot error-correction~\cite{Bombin2015b}, meaning that the reconstruction of stabilizer values from a measurement of the gauge generators is fault-tolerant to a sufficiently low rate of measurement error, even without repeated rounds of measurement. This may have practical utility for implementation of the prethermalization construction, specifically for the fault tolerant measurement of stabilizers at the end of an evolution to check that the symmetry has not been broken.
    
    A further possibility are fracton topological codes~\cite{Haah2011,Vijay2016,Williamson2016,Dua2019} which have unconventional symmetries that lead to a number of global relations that can scale in between a constant and subextensively with linear system size, depending on boundary conditions. With open boundary conditions these would-be relations instead generate global stabilizer symmetries, and associated error detecting codes, that are suitable for the application of the prethermalization construction. 
    In addition, there are subsystem fracton codes obtained by fractalizing two dimensional subsystem codes~\cite{Devakul2020b}, which support nontrivial gauge degrees of freedom that evolve even under perfect symmetry respecting local dynamics, while maintaining a number of logical qubits that can grow subextensively. 
\bibliography{refs_zero}
\end{document}